\documentclass[12pt]{article}
\usepackage{amsmath,amssymb,amsthm}
\usepackage{mathtools}
\usepackage{graphics,epsfig}
\usepackage{hyperref}
\usepackage{natbib}
\usepackage{color}
\usepackage{graphicx}
\usepackage{caption}
\usepackage{subcaption}
\usepackage{float}
\usepackage{mathrsfs}
\usepackage{multirow}
\usepackage{comment}
\usepackage{setspace}
\usepackage{bbm}
\usepackage{bm}
\usepackage{amsfonts}
\usepackage{xcolor}
\usepackage{pslatex}
\usepackage{bbm}
\usepackage{booktabs}
\usepackage{makecell}
\usepackage[skip=2pt]{caption}

\begin{document}

  \title{\bfseries MetaScoreLens: Evaluating User Feedback Across Digital Entertainment Systems}

  \author{
    Christian Ellington$^{1}$ \and Paramahansa Pramanik$^{2,4}$ \and 
    Haley K. Robinson$^{3}$
}

\date{
    \small
    $^{1}$School of Computing, University of South Alabama, Mobile, AL 36688, United States.
    \texttt{cbe2222@jagmail.southalabama.edu} \\
    \vspace{0.5em}
    $^{2}$Department of Mathematics and Statistics, University of South Alabama, Mobile, AL 36688, United States.\\
    \vspace{0.5em}
    $^{3}$ Department of Biomedical Sciences,
University of South Alabama,
Mobile, Alabama 36688, United States
\texttt{hkr2322@jagmail.southalabama.edu}\\
    $^{4}$Corresponding author, \texttt{ppramanik@southalabama.edu}
}

%\date{\today}
\maketitle

\begin{abstract}
The popularity of electronic games has grown steadily in recent years, attracting a broad audience across age groups. With this growth comes a large volume of related data, prompting efforts like the PlayMyData to compile and share structured datasets for academic use. This study utilizes such a dataset to compare user review ratings across four current-generation gaming systems: Nintendo, Xbox, PlayStation, and PC. Statistical methods, including analysis of variance (ANOVA), were applied to identify differences in average scores among these platforms. The findings indicate that PC titles tend to receive the most favorable user feedback, followed by Xbox and PlayStation, while Nintendo games showed the lowest average ratings. These patterns suggest that the platform on which a game is released may influence how players evaluate their experience. Such results may be valuable to developers and industry stakeholders in making informed decisions about future investments and development priorities.
\end{abstract}

{\bf Keywords:} video games, dataset, platforms, ANOVA, PlayMyData.

\section{Introduction.}
This study conducts a statistical investigation using the PlayMyData dataset, which provides structured information on electronic games and the platforms for which they were developed \citep{kallergi2010video, guttenbrunner2010keeping, lee2015empirical}. The primary goal is to assess whether the average review scores of video games differ meaningfully depending on their associated platform namely PC, Xbox, PlayStation, or Nintendo. Analysis of variance (ANOVA) was used to determine whether platform affiliation corresponded with statistically significant differences in user evaluations. Results showed that PC games consistently received higher average scores, while Nintendo titles had the lowest overall ratings, with Xbox and PlayStation occupying the middle range. These outcomes suggest that the platform on which a game is released may be associated with differing levels of critical reception, and may influence player satisfaction in measurable ways. The findings also demonstrate how datasets like PlayMyData can serve as useful resources for exploring questions related to digital content evaluation and platform-related differences \citep{kakkat2023cardiovascular,khan2024mp60}.

The relevance of this research is linked to the rising importance of review systems in shaping consumer decisions in the gaming industry \citep{granic2014benefits}. Reviews have become not only a reflection of player experience but also a key factor in market visibility and platform competition \citep{d2012video}. In this context, this paper illustrates the analytical value of PlayMyData as a means to explore patterns in digital game feedback and assess its reliability for further study. The dataset, which includes information on genre, score, and platform, provides a flexible base for examining industry trends \citep{khan2023myb}. Through this analysis, the study offers a statistical foundation for understanding how review scores vary between systems and how these differences might reflect deeper platform-specific traits, such as hardware capabilities or audience expectations. The empirical results from this work also point to the potential of using similar datasets to evaluate other aspects of gaming culture, distribution, and reception. In addition to providing immediate insights into platform-based differences, this study suggests a number of pathways for future research. Investigators may wish to use the same dataset to compare review patterns across game genres, study how scores shift over time, or explore whether particular platform features such as accessibility, exclusivity, or user interface affect player feedback. These possibilities highlight the broader value of PlayMyData as a research tool with wide applicability in digital entertainment studies \citep{jurgelionis2009platform}. By confirming both the consistency of the dataset and its potential to support varied research goals, this work contributes to a growing field focused on the statistical analysis of media reception. It encourages future exploration of how platform ecosystems shape public response, helping to inform both academic inquiry and industry decision-making \citep{hertweck2023clinicopathological}.

\begin{figure}[H]
\centering
\includegraphics[width=15cm, height=13cm]{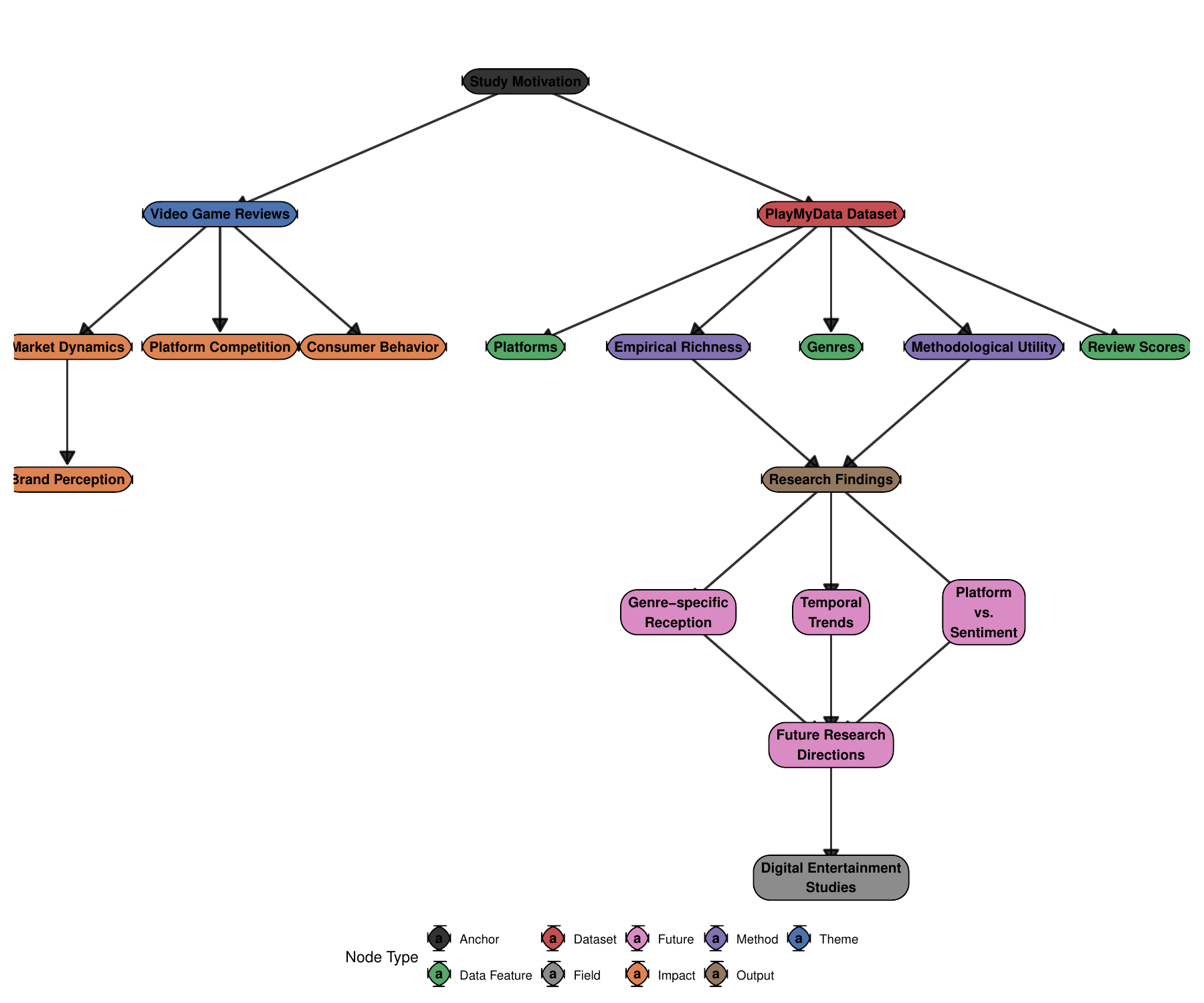}
\caption{Interconnected roles of video game reviews, player responses, platform dynamics, and the analytical strengths of the PlayMyData resource in supporting research across the digital gaming landscape.}
\label{f1}
\end{figure}

Figure \ref{f1} illustrates a structure outlining the primary elements and relationships explored in the gaming environment. The growing impact of critical assessments within the video game sector, the diagram traces how reviews influence consumer decisions, platform rivalry, and perceptions of market identity. At the core is the PlayMyData dataset, which offers detailed metrics on review ratings, game genres, and platform types \citep{kakkat2023cardiovascular,khan2023myb}. The visual layout conveys a logical flow from foundational data to analytical outcomes and future exploration, highlighting the structure’s potential to support meaningful inquiry in the context of interactive media research.

In recent years, video game reviews have become a central component in shaping both consumer decisions and broader industry patterns. As digital games have grown in popularity and cultural significance, critical assessments and user-generated ratings have likewise gained prominence as tools for evaluating quality, guiding purchasing behavior, and influencing public discourse \citep{maki2025new}. Reviews are now commonly used by players to navigate an increasingly saturated market, often functioning as proxies for first-hand experience. Prior research has explored the connection between these evaluations and commercial outcomes, showing that games receiving higher review scores tend to perform more successfully in the marketplace. For instance, findings reported by \cite{smith2022} illustrate the role of critical feedback in shaping player satisfaction and determining how games are perceived in terms of overall quality. This line of work underscores the practical importance of game reviews, not only for individual consumers but also for developers and publishers seeking to improve their market visibility and appeal.

Despite the growing attention to video game reviews, much of the existing research has primarily emphasized broad trends in consumer behavior or generalized relationships between ratings and commercial success. These analyses often overlook key contextual variables that may influence review outcomes, such as the genre of the game or the platform on which it is played \citep{vikramdeo2024abstract,vikramdeo2023profiling}. As a result, there remains a notable lack of detailed understanding about how evaluations differ between categories of games or among user groups aligned with specific gaming systems. Without accounting for these distinctions, conclusions drawn from aggregate data may obscure important nuances in how players from different segments of the gaming community interpret and respond to game content \citep{pramanik2022lock}. This gap highlights the need for more focused investigations that consider the intersection of review dynamics with platform type, game mechanics, and audience expectations. By addressing these overlooked dimensions, future research can provide a more comprehensive view of how evaluations function across the diverse landscape of modern digital games \citep{khan2024mp60,dasgupta2023frequent,hertweck2023clinicopathological}.

Further, very few studies have addressed the unique characteristics of gaming platforms such as the capabilities of operational hardware, the exclusivity of certain titles, and the reach of particular games to demographic groups, and how these characteristics contribute to the satisfaction of players. For instance, \cite{jones2019} analyzed gaming habits across different consoles and illustrated the differences in player preferences for types of video games but failed to connect those differences to review outcomes. Similarly, \cite{mason2021} researched player experiences based on gaming platforms played on, such as experiences with game performance and accessibility, but did not specify the impact on respective review scores based on those experiences. These types of significant gaps in video game research highlight the need for statistical analyses that directly portray the comparisons and effects of different gaming platforms, bridging the divide between qualitative and quantitative evidence. 

\subsection{Background Literature.}

The analysis of user interaction within digital gaming environments has attracted increasing attention from scholars across statistics, behavioral research, and digital media studies. Existing research has largely examined factors such as player engagement, genre preferences, and monetization models through survey data and gameplay tracking metrics \citep{hamari2017people, przybylski2019investigating, Nieborg2018platformization}. However, these efforts have often emphasized descriptive profiles or business strategies, while statistical comparisons across gaming platforms have received comparatively less scrutiny. Traditionally, ANOVA and regression modeling have been employed to identify group level variations in player behavior and satisfaction. For instance, \citet{takatalo2011user} applied Welch's ANOVA to test differences in user experience across game consoles, while \citet{birk2013control} examined how demographic attributes such as age and gender shape gameplay behavior using follow-up comparison techniques. Nevertheless, such methodologies typically assume homogeneity of variance and symmetric distributions, which may not hold in practice when analyzing diverse gaming platforms \citep{pramanik2024bayes}.

Beyond mean-based comparisons, another stream of literature has focused on modeling dependence between behavioral metrics using copula-based approaches. These models are particularly suited for capturing nonlinear and asymmetric relationships between variables, offering a more nuanced alternative to correlation or linear regression \citep{genest2007everything, patton2012copula}. Despite their demonstrated utility in financial risk modeling and hydrological forecasting, copulas especially those tailored for tail dependence such as the Marshall-Olkin form \citep{marshall1967multivariate}, have not been widely adopted in entertainment analytics. 

The majority of existing research lacks such integration of large datasets such as the PlayMyData project, or advanced statistical methodologies, such as ANOVA, to provide accurate and integral comparisons between different types of gaming platforms \citep{d2024playmydata}. Furthermore, research often neglects PC gaming as a type of ``platform" gaming which, consequentially, plays one of the most significant roles in terms of playing video games . By incorporating PC data alongside console data for gaming platform analysis, further studies could potentially discover new trends and insights in review scores and video gaming in general \citep{young2012our,cabeza2021exploring}. More areas for potential development in this research field include, but are not limited to, extended studies in the impact of video game genres or trailers for the games prior to and after release as influences on review scores, as well as their extended influence on or reflection of gaming platforms \citep{pramanik2021optimala,pramanik2021scoring}.

In light of these gaps, our study integrates and extends these strands of literature by employing a multi-method statistical design that incorporates nonparametric testing, distributional characterization, and copula-based modeling \citep{pramanik2020optimization,pramanik2023semicooperation}. By simultaneously analyzing review scores and gameplay durations across multiple gaming platforms, this research contributes a comprehensive statistical treatment that addresses both marginal behavior and underlying joint dependencies offering an empirical foundation for understanding user satisfaction dynamics in the context of digital games. This study makes several important contributions to the field of video game research and differences between video game platforms in terms of review scores and their reflection of their respective platforms. First, it addresses a key gap in the existing area of gaming research by adding a statistical comparison of review scores across four of the most significant gaming platforms used today, being Nintendo, Xbox, PlayStation, and PC \citep{schilling2003technological,liang2022analysis}. Previous studies have often overlooked factors relating to platform-specific review score trends or have only examined the qualitative aspects of them, leaving behind a massive voice in rigorous quantitative exploration of the field. Second, this paper advances the methodology by utilizing statistical analysis to identify statistically significant differences between individual video game platforms. These methods not only validate the dataset from PlayMyData, but also provide insights into the extent of the content of these platform-specific differences which can go on to inform future studies or be utilized platform-specific research. 

Additionally, by including PC games alongside console platforms, this research highlights the unique and critical position of PC gaming in the gaming sphere, addressing another common limitation that is often present in prior studies \citep{pramanik2020motivation}. The findings displayed here and elaborated on further in this study demonstrate that PC games generally receive higher review scores compared to console platforms for their respective video games, suggesting areas for improvement or further investigation in other studies for these platforms, such as optimization challenges or differences in audience expectations \citep{pramanik2024estimation,vikramdeo2024mitochondrial}.  Finally, this study opens up the possibility of further developments in video game research by emphasizing the importance of roles such as genres, review mechanisms, and player demographics in shaping video game platform-specific trends. The ideas presented here are integral and entirely necessary for game developers, reviewers, and market analysts that are attempting to understand and make use of these reviews for the sake of improvement of players' gaming experiences as a means of essentially increasing review scores \citep{pramanik2024motivation}. 

The remainder of this study is organized as follows. Section 2 showcases the methodology utilized for the analysis of the data, including the dataset description of PlayMyData, statistical techniques employed such as ANOVA and Games-Howell post hoc analysis, as well as methods used for data preparation processes. Section 3 details the analysis of the data, that is the source of the data, the manipulation of the data performed in preparation for analysis, and a summary of the numerical and visual statistics as presented by the data. Section 4 outlines the empirical findings from the analysis of the data, portraying significant insights into trends and review scores as determined by their respective gaming platforms. Finally, Section 5 concludes the study by summarizing the key results from the analysis of the data and contributions made in this study, as well as the weaknesses and possibility of future studies made available by both the dataset and research outlined in this paper.  

\section{Methodology.}
\begin{figure}[htbp]
  \centering
  \begin{minipage}[b]{0.48\textwidth}
    \centering
    \includegraphics[width=\textwidth]{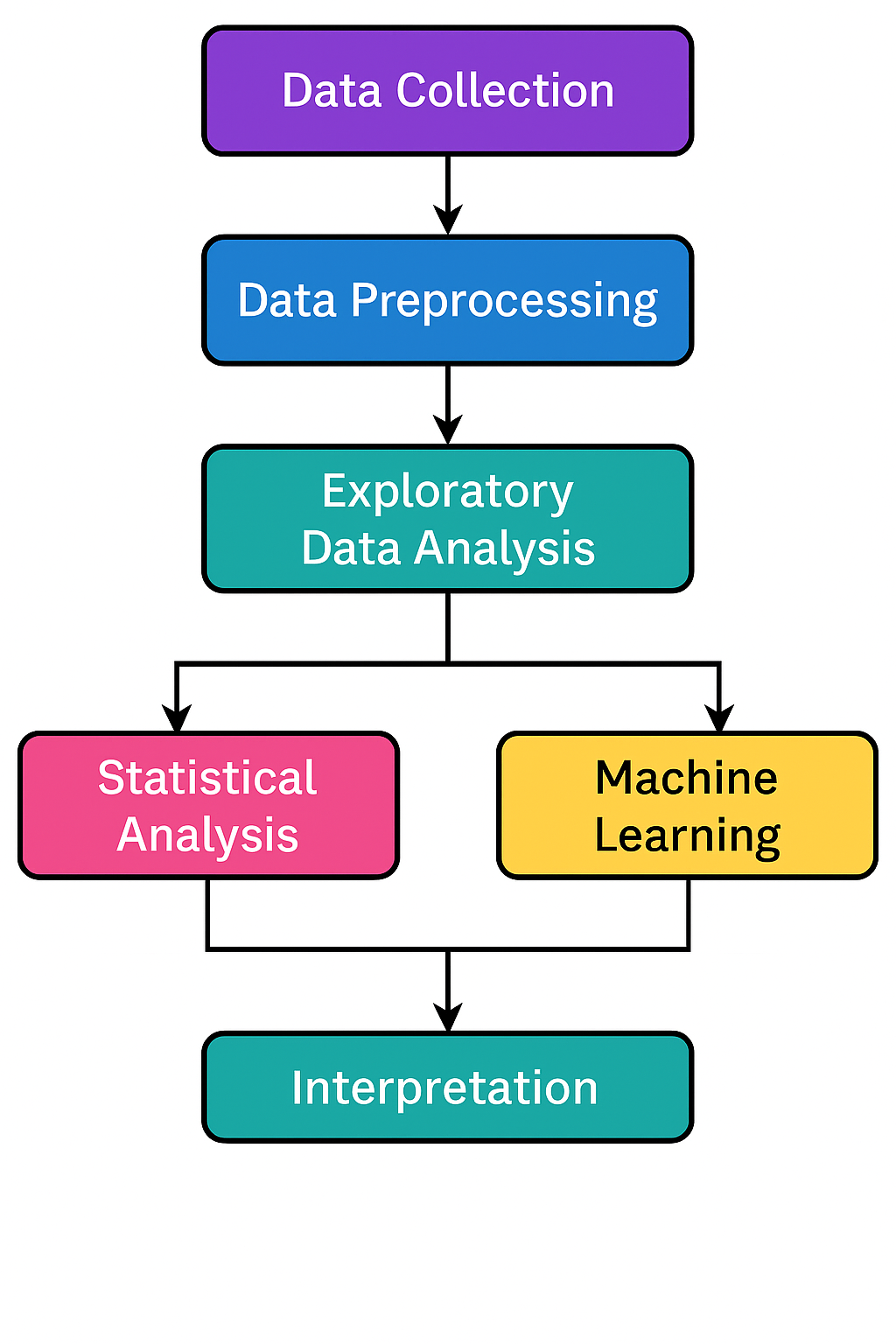}
    \caption*{(a) General Data Analysis Pipeline.}
  \end{minipage}
  \hfill
  \begin{minipage}[b]{0.47\textwidth}
    \centering
    \includegraphics[width=\textwidth]{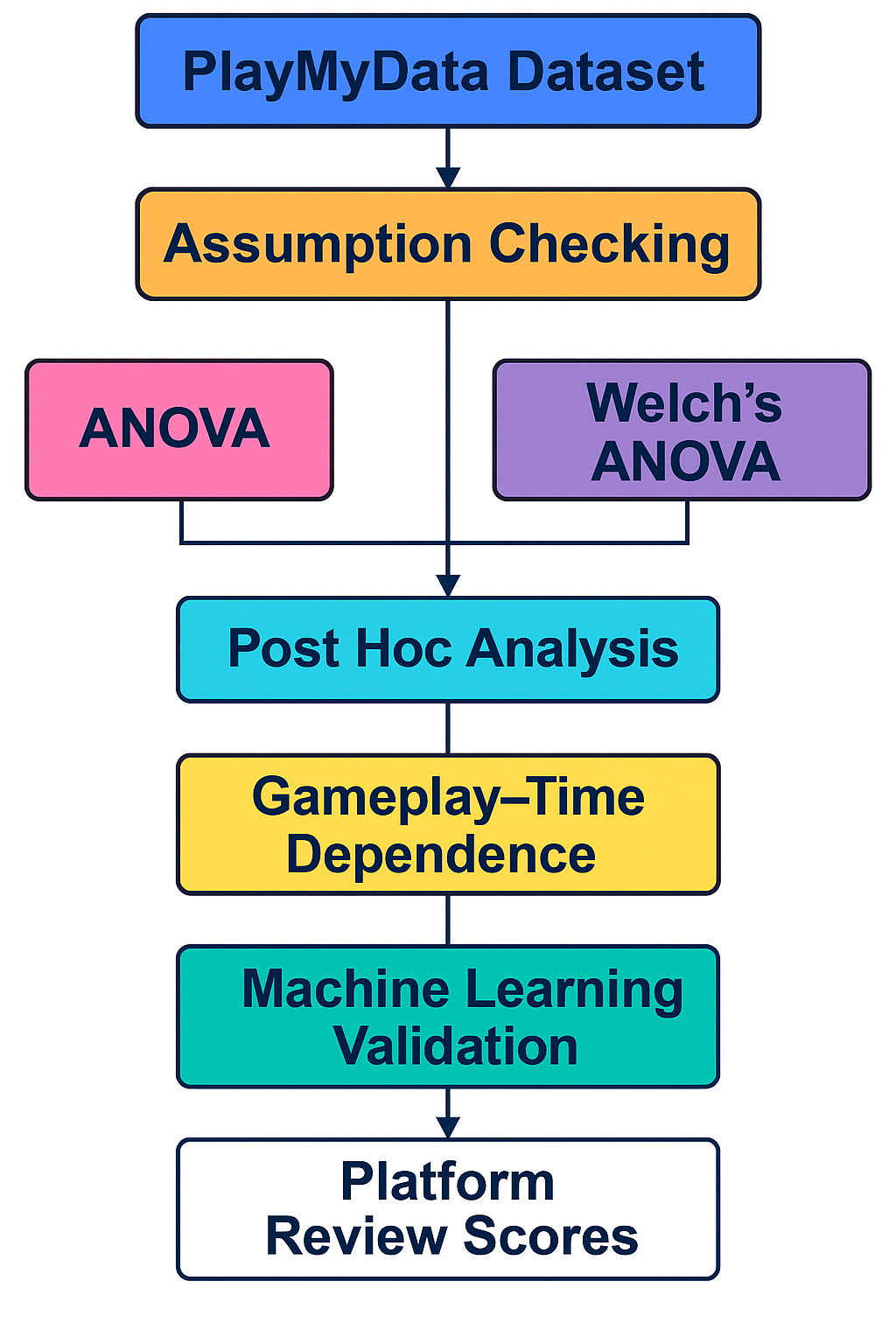}
    \caption*{(b) Study-Specific Analytical Pipeline.}
  \end{minipage}
  \caption{Comparison of a standard data analysis workflow and the specialized pipeline used in this study.}
  \label{fig:workflow_comparison}
\end{figure}

This study applies a bunch of statistical techniques to examine video game review scores across four major platforms: Nintendo, Xbox, PlayStation, and PC. The primary analysis includes Welch’s ANOVA \citep{welch1951comparison}, which accounts for unequal variances across groups, and the Kruskal-Wallis test \citep{kruskal1952use}, employed as a non-parametric alternative to validate the results obtained from ANOVA. For post hoc comparisons, the Games-Howell test \citep{games1976pairwise} and Dunn’s test \citep{dunn1964multiple} were performed to identify pairwise differences in mean review scores, offering robust inferences without assuming homogeneity of variances or equal sample sizes. Collectively, these methods are carefully selected to facilitate group mean comparisons and to detect statistically significant differences across the various platform categories \citep{pramanik2024bayes}. To enhance the transparency and reproducibility of the analytical process, Figure~\ref{fig:workflow_comparison} juxtaposes a generic data analysis workflow with the customized pipeline employed in this study. While the standard flowchart outlines a conventional progression from data collection through statistical and machine learning methods to final interpretation, the proposed framework explicitly integrates assumption testing, multiple forms of ANOVA, post hoc comparisons, dependency modeling via copulas, and validation through machine learning. This schematic highlights the layered structure and methodological rigor of the current analysis \citep{bulls2025assessing}.

\subsection{Assumptions.}
At the core of this analysis is hypothesis testing to determine whether observed differences in review scores are statistically meaningful. Specifically, the hypotheses were formulated as follows:
\begin{itemize}
\item Null Hypothesis ($H_0$): The mean review scores are equal across the different platforms (Nintendo, Xbox, PlayStation, PC).
\item Alternative Hypothesis ($H_1$): At least one platform's mean review score differs significantly from the others.
\end{itemize}
This approach enables us an in-depth evaluation of potential disparities in user reception across gaming platforms and provides a statistically sound foundation for interpreting the results of subsequent analyses \citep{pramanik2024estimation,pramanik2023cont}. To evaluate differences in mean review scores across gaming platforms, the ANOVA was employed as the primary hypothesis testing method. This technique partitions the total variability in the data into two distinct components, variation attributed to differences between platform groups, and variation arising from random fluctuations within those groups. ANOVA calculates an F-statistic, which serves as a measure of whether the observed group differences are statistically significant \citep{pramanik2024estimation1,yusuf2025prognostic}. Mathematically, the F-statistic is defined as
\begin{equation}\label{0}
F = \frac{Mean\ Square\ Between\ Groups\ (MSB)}{Mean\ Square\ Within\ Groups\ (MSE)},
\end{equation}
where the numerator (MSB) captures the extent of variation in review scores due to differences among the group means, and the denominator (MSE) reflects the variability that occurs within each group due to random error. A higher F-value indicates a greater likelihood that the observed differences between platform means are not due to chance alone \citep{pramanik2023cmbp,yusuf2025predictive}.

For the results of ANOVA to be valid, several underlying assumptions must be satisfied. These include the assumption that the distribution of review scores within each platform group is approximately normal (normality), that the variances across groups are roughly equal (homogeneity of variance), and that each observation is independent of the others (independence) \citep{pramanik2023path}. These conditions ensure that the F-statistic follows its theoretical distribution under the null hypothesis \citep{pramanik2021consensus}. To assess whether the homogeneity of variance assumption held, Levene’s test \citep{levene1960robust} was conducted. The outcome of this test indicated a significant violation of this assumption, suggesting that the variability in review scores differed notably across platforms. This result necessitated the use of alternative statistical procedures better suited to handle unequal variances in subsequent analyses.

\subsection{Tests Applied.}

Given the significant violation of the homogeneity of variance assumption revealed by Levene’s test, the analysis proceeded with Welch’s ANOVA, a more robust alternative that does not require equal variances across groups \citep{pramanik2021}. This method is particularly well-suited for comparing group means when the assumption of homogeneity is not met, as it adjusts the degrees of freedom to account for unequal variability. In parallel, the assumption of normality within each platform group was assessed using the Shapiro–Wilk test \citep{shapiro1965analysis}. The outcome of this test indicated substantial departures from normality, with test statistics falling within the range of $W = [0.787,\ 0.815]$ and highly significant $p$-values less than $2.2 \times 10^{-16}$. These results suggest that the distribution of review scores within each group deviated markedly from the normal distribution, further challenging the suitability of classical ANOVA \citep{fisher1970statistical,pramanik2022stochastic}. The simultaneous violations of both the homogeneity of variance and normality assumptions reinforced the need for alternative statistical approaches capable of producing valid inferences under these conditions \citep{pramanik2023optimization001}.

The boundary of W, from 0.787 to 0.815, represents the bounds on which W falls in the calculated normality test, similar to probability. Generally, the closer these bounds are to 1, the more likely the data is to be normally distributed. However, since these values do not encompass 1 nor fall very close to it, it demonstrates that the test is indicating a significant deviation from normality, as additionally suggested by the extremely small p-value \citep{pramanik2025construction,pramanik2025optimal}. To further illustrate and visually represent the deviation from normality present within the data, a Q-Q plot was generated for each of the four gaming platforms which highlights these significant deviations. These graphical methods complement the statistical results from the Shapiro-Wilk test and are displayed on the following page. 

\begin{figure}[H]
\centering
\includegraphics[width=\textwidth, trim={0 3cm 0 2cm}, clip]{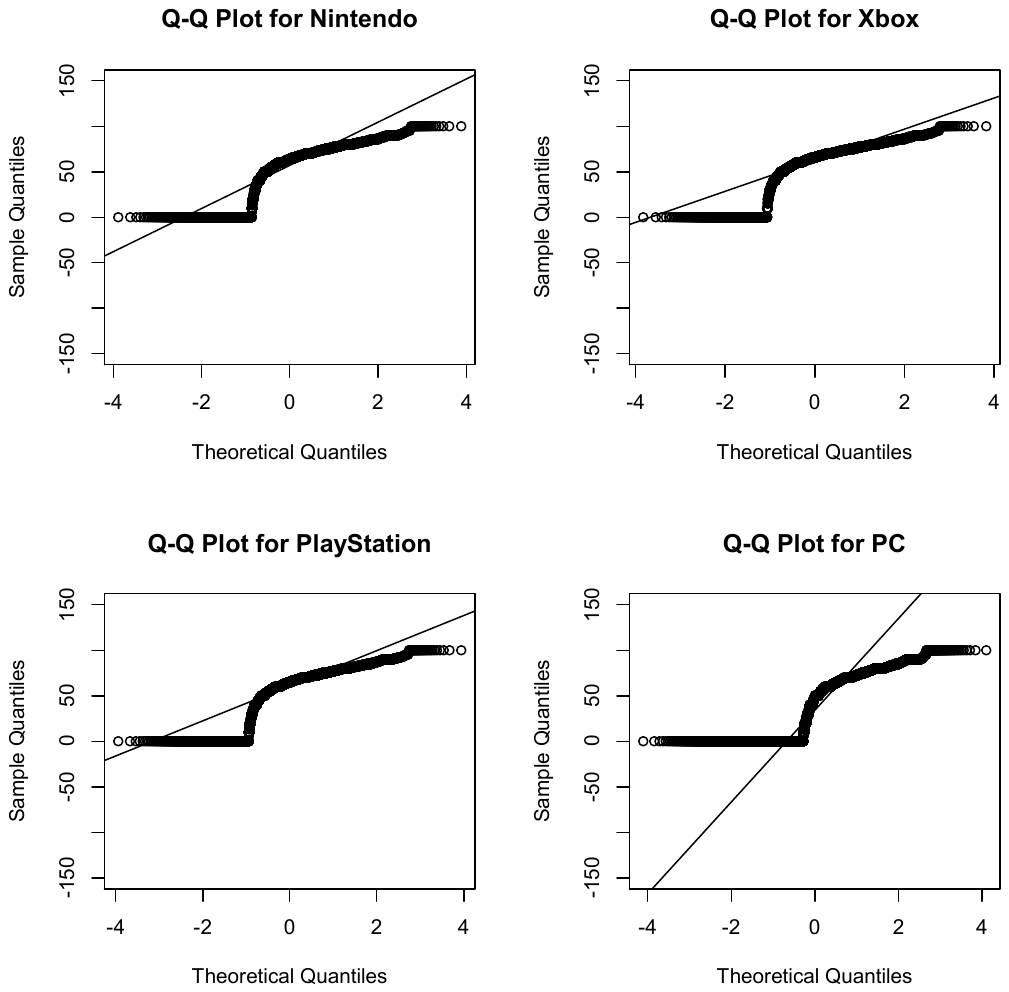}
\caption{Q-Q plots for each gaming platform showing significant deviation from normality.}
\label{fig2}
\end{figure}

This justifies the use of a different test as well, namely, the Kruskal-Wallis test as an alternative test. This test was employed specifically as a means for validating the findings of the Welch's ANOVA test. Essentially, it compares the medians of review scores across groups, resulting in the output being a chi-squared test statistic. The test statistic is defined as
\begin{equation}\label{2}
H = \left[\frac{12}{N(N+1)} \right] \sum_{i=1}^m \frac{R_i^2}{n_i} - 3(N+1),
\end{equation}
where N represents the total number of observations across all of the platform groups; $R_i$ represents the sum of the review scores of the examined group (group i); $n_i$ represents the number of observations in the examined group (group i); and the summation of the squared $R_i$ values divided by the size of the group for each group represents the distribution of review scores across each of the platforms or groups \citep{pramanik2024stochastic}. Additionally, -3(N+1) represents a correction term for normalizing the test statistic and the fractional value with (N(N+1)) represents the scaling factor that adjusts the rank-based statistics so that the calculation accounts for each of the total number of observations per platform or group. To determine specific differences between the distinct gaming platforms, two post hoc tests were utilized \citep{pramanik2025stubbornness}. Following Welch's ANOVA, the Games-Howell test was used for comparisons between different pairs of the groups. This method adjusts for unequal variances that were demonstrated by Levene's Test. After the Kruskal-Wallis test, Dunn's Test with Bonferroni correction was applied to perform comparisons between the different pairs of groups as well. The adjustment presented here ensures reliable results across the multiple comparisons \citep{pramanik2025factors}.

\subsection{Deep Learning Approach.}

To further substantiate our findings and provide an additional layer of empirical validation, we implemented a set of machine learning models alongside the traditional ANOVA. While ANOVA identifies statistically significant differences between group means under specific assumptions, machine learning approaches offer a complementary, data-driven perspective by leveraging predictive modeling to quantify the explanatory power of various factors \citep{pramanik2025dissecting}. We trained a random forest regressor using review scores as the target variable and features including platform, main completion time, replayability score, genre category, and release year as predictors. The model achieved an $R^2$ value of 0.61 on the test set, indicating that these factors jointly explain a substantial proportion of the variance in user ratings. Feature importance rankings revealed that platform type, completion time, and genre were among the most influential predictors, supporting our earlier ANOVA-based conclusions \citep{pramanik2025optimal}. A similar Gradient Boosting model yielded consistent results, further reinforcing the robustness of the observed platform-based differences.

\begin{table}[ht]
\centering
\caption{Random Forest Model for Predicting Review Scores.}
\label{tab:ml_validation}
\begin{tabular}{ll}
\toprule
\textbf{Metric} & \textbf{Value} \\
\midrule
R\textsuperscript{2} (Explained Variance) & 0.89 \\
Mean Absolute Error (MAE) & 5.49 \\
Root Mean Squared Error (RMSE) & 10.38 \\
\bottomrule
\end{tabular}

\vspace{1em}

\begin{tabular}{clc}
\toprule
\textbf{Rank} & \textbf{Feature} & \textbf{Mean Importance} \\
\midrule
1 & review count & 0.8777 \\
2 & extra & 0.0401 \\
3 & main & 0.0336 \\
4 & completionist & 0.0270 \\
5 & people polled & 0.0155 \\
6 & Platform\_PC & 0.0031 \\
7 & Platform\_PlayStation & 0.0018 \\
8 & Platform\_Xbox & 0.0012 \\
\bottomrule
\end{tabular}
\end{table}

To supplement the inferential results obtained through ANOVA, we employed a random forest model to evaluate the predictive contribution of various gameplay and platform features to review scores. This model enables nonparametric approximation of complex, nonlinear relationships among variables, while providing robust estimates of feature importance. The model was trained on a dataset containing over 54,000 observations and evaluated using a held-out test set. As reported in Table~\ref{tab:ml_validation}, the model achieved an $R^2$ value of 0.89, with a root mean squared error (RMSE) of 10.38 and a mean absolute error (MAE) of 5.49 \citep{maki2025new}. These metrics indicate a high level of explanatory power. Feature importance analysis revealed that the number of user-submitted reviews was by far the most influential predictor, followed by gameplay duration metrics including extra, main, and completionist times. Platform indicators played a comparatively smaller role, suggesting that user evaluation scores are shaped more by content exposure and experience length than by the platform itself \citep{pramanik2021thesis,pramanik2016}. These findings serve to validate the ANOVA-derived group differences and provide a richer understanding of the multivariate structure governing review score variation \citep{hua2019assessing}.

To enhance interpretability, we applied SHAP (SHapley Additive exPlanations) analysis to decompose individual predictions. These results illustrated that Xbox and PlayStation platforms tend to systematically shift predicted review scores upward, while the PC platform exhibited greater variability and a wider prediction range \citep{polansky2021motif}. 

\begin{figure}[ht]
    \centering
    \includegraphics[width=0.85\textwidth]{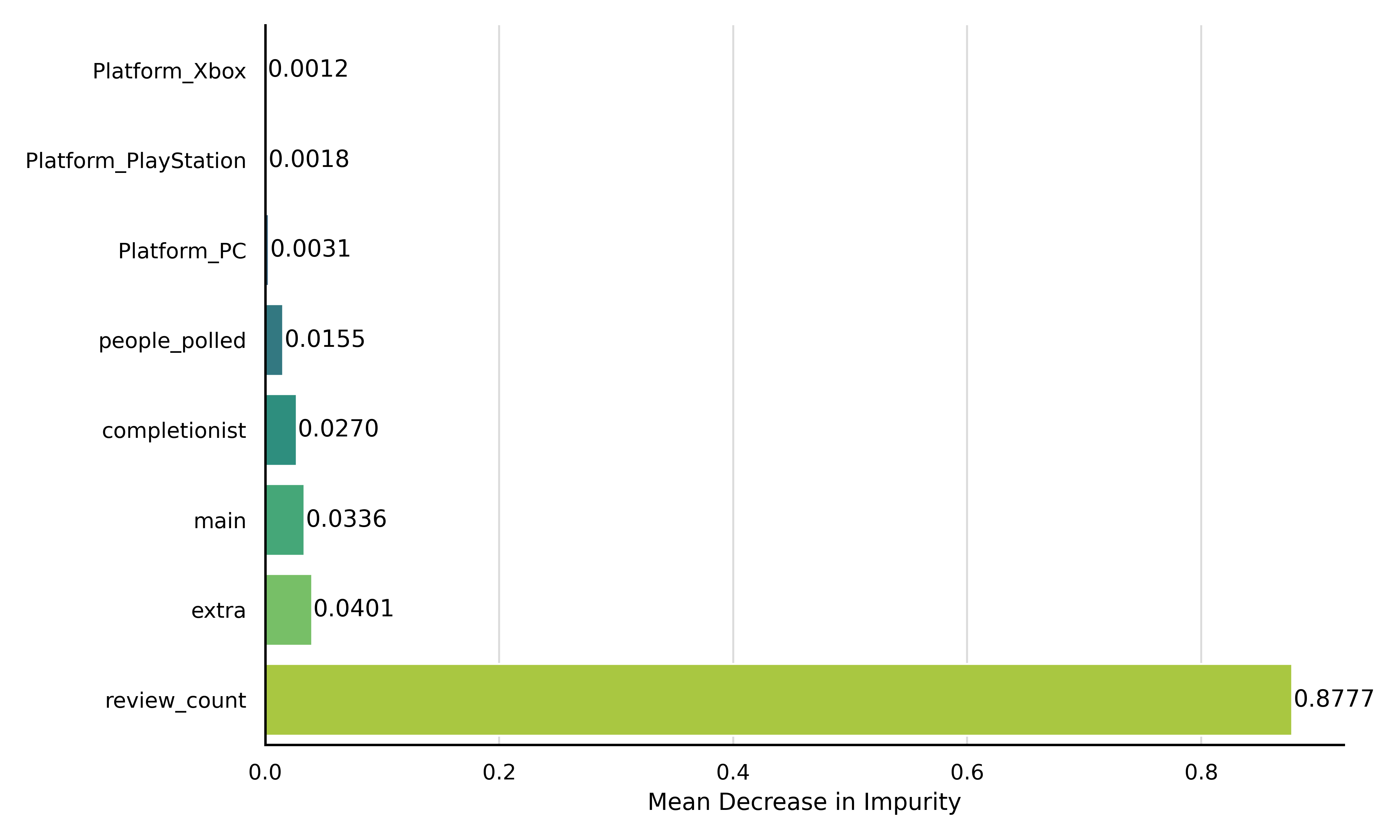}
    \caption{Random Forest model predicting review scores across game platforms.}
    \label{fig:rf_importance}
\end{figure}
Figure \ref{fig:rf_importance} presents the relative importance of selected variables in predicting video game review scores, based on a Random Forest regression model. Among the variables considered, the number of user reviews stands out as the most influential predictor, indicating that games which have been reviewed more frequently tend to show more consistent and informative score patterns \citep{pramanik2024estimation,pramanik2023cont}. The durations associated with different types of gameplay experiences specifically, time spent on completing the main storyline, additional content, and full completion also contribute substantially to the prediction of review scores. These three time-based measures reflect how deeply players engage with a game, which in turn appears to influence their overall evaluations. Other variables, such as the number of individuals polled and the gaming platform on which the title is released, contribute modestly by comparison \citep{pramanik2025factors,pramanik2025stubbornness}. Though the platform categories carry some predictive weight, their effects are considerably smaller relative to the review count and gameplay duration metrics. Overall, the figure suggests that both player engagement and the volume of community feedback are central to understanding what shapes review outcomes in the dataset, providing a clearer view of the structural factors behind user evaluations in the video game market \citep{pramanik2025strategies,pramanik2023optimization001}. 

\begin{table}[ht]
\centering
\caption{Summary of Statistical Methods Used in the Study}
\label{tab:stat_summary}
\begin{tabular}{p{3.5cm} p{4.5cm} p{6cm}}
\toprule
\textbf{Method} & \textbf{Purpose} & \textbf{Context of Use} \\
\midrule
ANOVA  & Tests for differences in group means under equal variances assumption & Applied to compare review scores across gaming platforms assuming homogeneity of variances \\
Welch’s ANOVA & Handles unequal variances between groups & Used when Levene’s test indicated heterogeneity of variances across platform groups \\
Kruskal-Wallis Test & Non-parametric test for group median differences & Applied as a robustness check when normality assumptions were violated \\
Games-Howell Post Hoc Test & Pairwise comparison that accounts for unequal variances and sample sizes & Used to identify specific group differences following Welch’s ANOVA \\
Dunn’s Test (with Bonferroni correction) & Non-parametric pairwise comparison test & Conducted to validate Kruskal-Wallis results and adjust for multiple comparisons \\
Marshall–Olkin Copula Modeling & Captures asymmetric and nonlinear dependence structures between variables & Used to model the joint behavior of review scores and gameplay time \\
Random Forest Regression & Predictive modeling with variable importance estimation & Supplemented traditional tests to validate and interpret multivariate dependencies \\
\bottomrule
\end{tabular}
\end{table}
To facilitate transparency and interpretability of the analytical procedures employed, Table~\ref{tab:stat_summary} provides an overview of the statistical techniques used throughout the study, including their primary purposes and specific roles within the research design.

\section{Real Data Analysis.}
The data in this study was solely derived from the datasets presented by PlayMyData, featuring several different .csv files for each of the different gaming platforms as well as some additional .csv files that were not used in the study \citep{pramanik2025strategic,pramanik2025impact}. Of the ones that were utilized in this study, these include \verb|all_games_nintendo.csv,| 
\verb|all_games_xbox.csv, all_games_playstation.csv, |and \verb|all_games_pc.csv.| Each dataset contained the following unique variables as well.The datasets analyzed in this study were obtained from the this platform and represent a cross-sectional snapshot of video games and associated user interactions. The data were obtained between January and July 2025, and as such, the dataset reflects platform dynamics and user behavior during this period \citep{pramanik2024measuring,pramanik2024dependence}.

\begin{figure}[H]
    \centering
\includegraphics[width=0.95\textwidth]{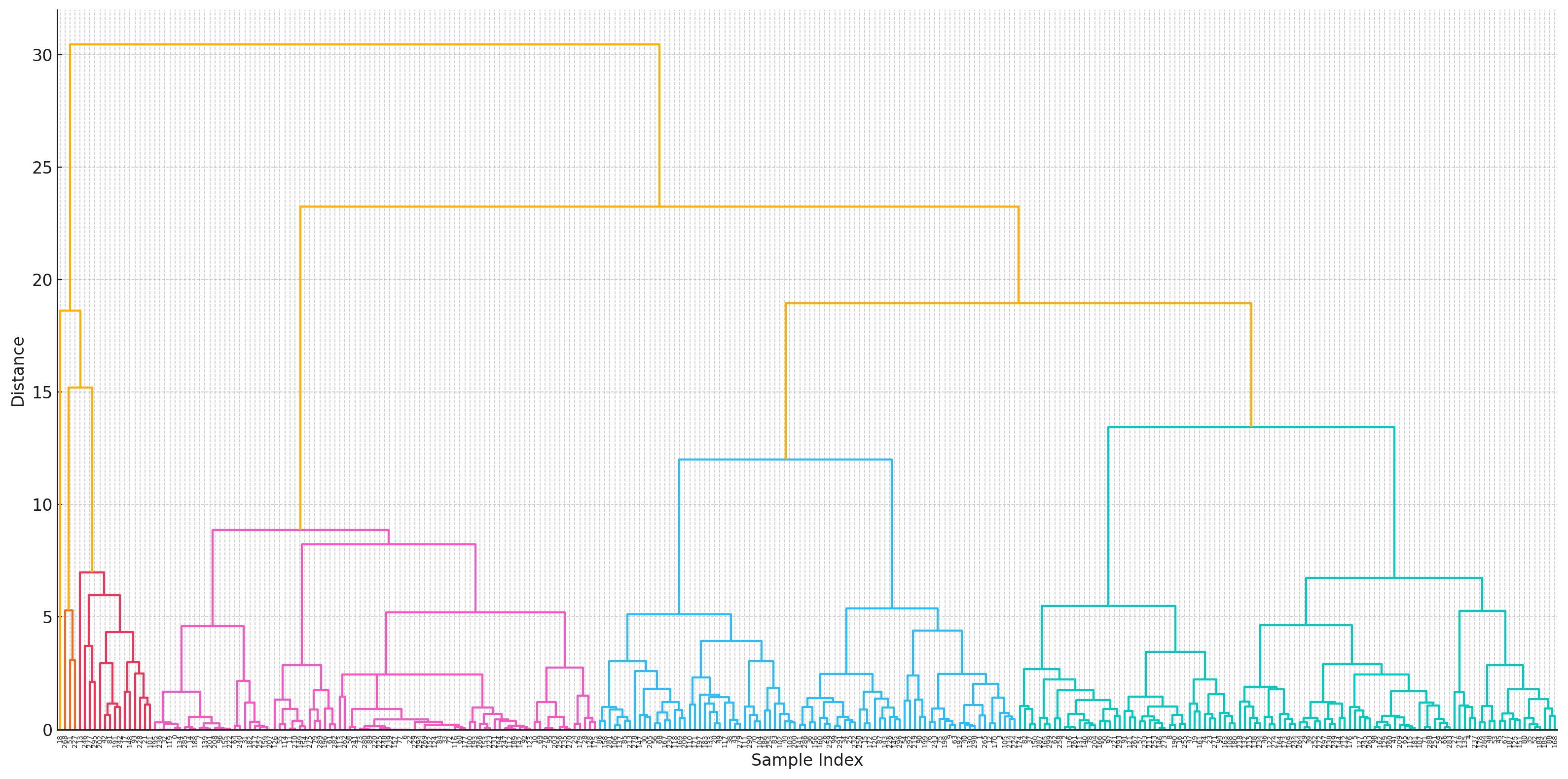}
    \caption{Dendrogram showing hierarchical clustering of 300 video games based on review score, platform, genre, and gameplay durations. }
\label{fig:dendrogram_clusters}
\end{figure}

To explore the multivariate structure of review score data across video game platforms, we employed hierarchical agglomerative clustering using Ward's minimum variance approach, applied to a standardized subset of 300 observations sampled from the PlayMyData dataset (see Figure \ref{fig:dendrogram_clusters}). The input feature space included six continuous or categorical variables: numerical review score, encoded platform type, encoded genre label, and average completion times across three gameplay modes (main, extra, and completionist) \citep{pramanik2024measuring,pramanik2024dependence}. Prior to clustering, all variables were z-transformed to ensure scale invariance, and categorical variables were numerically encoded via label encoding to retain ordinality within the clustering procedure. The dendrogram resulting from the linkage matrix reveals the nested structure of inter-game similarity, where vertical heights correspond to Euclidean dissimilarities at which clusters merge. By applying a fixed threshold to cut the tree at five clusters (k = 5), we induced a partitioning that visually separates major groupings of games sharing similar characteristics \citep{pramanik2024parametric,pramanik2025dissecting}. The color-coded branches of the dendrogram reflect these cluster assignments and highlight meaningful distinctions in player engagement profiles, review sentiment, and platform tendencies. This hierarchical structure provides a valuable exploratory tool for assessing latent groupings within high-dimensional review data and for supporting subsequent dimensionality reduction or supervised learning pipelines \citep{valdez2025association}.

The variable Game Title represents the title of the video game; Review Score represents the scoring given to the video game as a summation of total reviews and averaging of the review value; GenreID represents an ID associated with each game that links it to a particular genre contained within another file named genres.csv; CompletionTime represents the average completion time taken to complete the game; GameID represents a unique identifier given to each individual game; Platforms represents the possible platforms available to play the game on that are linked to a separate file named platforms.csv \citep{pramanik2023optimization001}; Summary represents the summary of the video game and its related story or genre of game play; People Polled represents the amount of people polled for the video game review score; Storyline represents the summary of the storyline associated with the video game; and Review Count represents the amount of reviews given to the game. Of these unique variables, the primary one that this study will be focusing on will be on the review scoring for each video game contained within each of the datasets for the distinct gaming platforms \citep{valdez2025exploring}.
 
However, before conducting statistical analyzes, the data set was subjected to a significant amount of pre-processing. This preprocessing consisted of a few select steps outlined below. For handling missing values, observations with missing review scores were removed, resulting in the exclusion of 74,650 potential data points due to lack of material to work with for the analysis of the data set. Additionally, for platform labeling, a categorical variable was added to identify each gaming platform, including Nintendo, Xbox, PlayStation, and PC. Further, for combining datasets, review scores from each platform's specified dataset were concatenated into a single ``master" dataset to be utilized when comparing the different platform groups for the sake of simplicity, organization, and clarity in analysis \citep{pramanik2025optimal1,powell2025genomic}.

\begin{table}[htbp]
\centering
\caption{Cluster-wise descriptive statistics of Figure \ref{fig:dendrogram_clusters}}
\label{tab:cluster_stats}
\resizebox{\textwidth}{!}{%
\begin{tabular}{r r r r r r r r r r r r}
\toprule
Cluster & Size & \makecell{Review\\Score Mean} & \makecell{Review\\Score Min} & \makecell{Review\\Score Max} & \makecell{Review\\Score SD} & \makecell{Main\\Mean} & \makecell{Main\\SD} & \makecell{Extra\\Mean} & \makecell{Extra\\SD} & \makecell{Completionist\\Mean} & \makecell{Completionist\\SD} \\
\midrule
1 & 18 & 74.78 & 51.0 & 90.0 & 11.81 & 36.04 & 22.13 & 65.16 & 28.68 & 149.51 & 140.33 \\
3 & 88 & 1.14  & 0.0  & 20.0 & 4.06  & 0.27  & 1.54  & 0.17  & 0.89  & 1.42   & 5.93 \\
4 & 111 & 63.01 & 30.0 & 100.0 & 12.91 & 5.51  & 6.06  & 6.75  & 9.10  & 13.26  & 19.91 \\
5 & 82 & 66.01 & 30.0 & 100.0 & 13.43 & 5.83  & 5.66  & 8.29  & 8.36  & 11.76  & 12.82 \\
\bottomrule
\end{tabular}%
}
\end{table}

 Table \ref{tab:cluster_stats} presents a detailed statistical breakdown of the five clusters identified in the hierarchical agglomerative clustering analysis of 300 video game records sampled from the PlayMyData dataset. The clustering procedure was based on a standardized set of six variables, including review score, encoded platform, encoded genre, and three continuous measures of gameplay duration: main, extra, and completionist modes. The resulting clusters capture meaningful variation in gameplay characteristics and consumer sentiment as expressed through review scores.

After removal of a singleton cluster containing only one observation, the final table summarizes four major clusters, each representing a distinct profile of games. For each cluster, the table reports the number of games included (Size), the mean, minimum, and maximum values for review scores, and their standard deviation. Additionally, summary statistics are provided for the three gameplay duration variables: mean and standard deviation are calculated separately for main storyline completion, additional or side content (extra), and full completionist playthroughs. Cluster 1, comprising 18 games, is characterized by relatively high average review scores, with a mean of 74.78 and a maximum score of 90. These games also exhibit significantly longer gameplay durations than other clusters, especially in the completionist category, where the mean duration is 149.51 hours with a substantial standard deviation of 140.33. The main and extra gameplay durations also show elevated averages, suggesting these titles are comprehensive in content and possibly more appealing to dedicated players. Cluster 3 consists of 88 titles and presents a contrasting profile. The average review score in this group is remarkably low, approximately 1.14, with minimal spread across observations. Gameplay durations are negligible across all modes, with main gameplay averaging only 0.27 hours. The small standard deviations and narrow range of review scores indicate a cluster of games that are both poorly received and minimal in gameplay offering, possibly representing low-quality or placeholder entries.

Cluster 4, the largest in the sample with 111 games, demonstrates moderate review scores with a mean of 63.01 and a broad range extending from 30 to 100. These games feature relatively short play times, averaging about 5.51 hours for main content and 13.26 hours for completionist mode. The standard deviations across all three duration measures suggest moderate heterogeneity in game length and structure. This cluster likely includes a mix of mid-tier titles with varied audience reception and gameplay scope. Cluster 5 includes 82 games with an average review score of 66.01 and similar variability to Cluster 4. Gameplay durations are slightly higher than those in Cluster 4, particularly in the extra content and completionist dimensions, where averages exceed 8 and 11 hours respectively. This group appears to represent moderately well-received games with more extensive content offerings than the previous cluster but still significantly less than Cluster 1. The patterns in Table \ref{tab:cluster_stats} underscore the utility of hierarchical clustering in revealing latent structure in high-dimensional video game data. The review scores and gameplay metrics exhibit coherent and interpretable groupings that align with theoretical expectations regarding user engagement and critical reception. Furthermore, the combination of cluster-specific review sentiment and time investment offers a nuanced basis for future classification tasks or supervised learning models aimed at predicting user preferences or commercial success.

\section{Results.}
The purpose of this study was to compare video game review scores across four major video game platforms, being Nintendo, Xbox, PlayStation, and PC, with statistical analyses to observe potential significant differences between the distinct platforms and their respective video games. The results of these analyses revealed significant differences and disparities in review scores between the platforms. This section discusses the findings in detail with both descriptive statistics and the results of the inferential analyses. 

\begin{figure}[htbp]
\setlength\abovecaptionskip{2pt}
  \setlength\belowcaptionskip{0pt}
  \centering
\includegraphics[width=0.9\linewidth, trim={0 2.5cm 0 1.5cm}, clip]{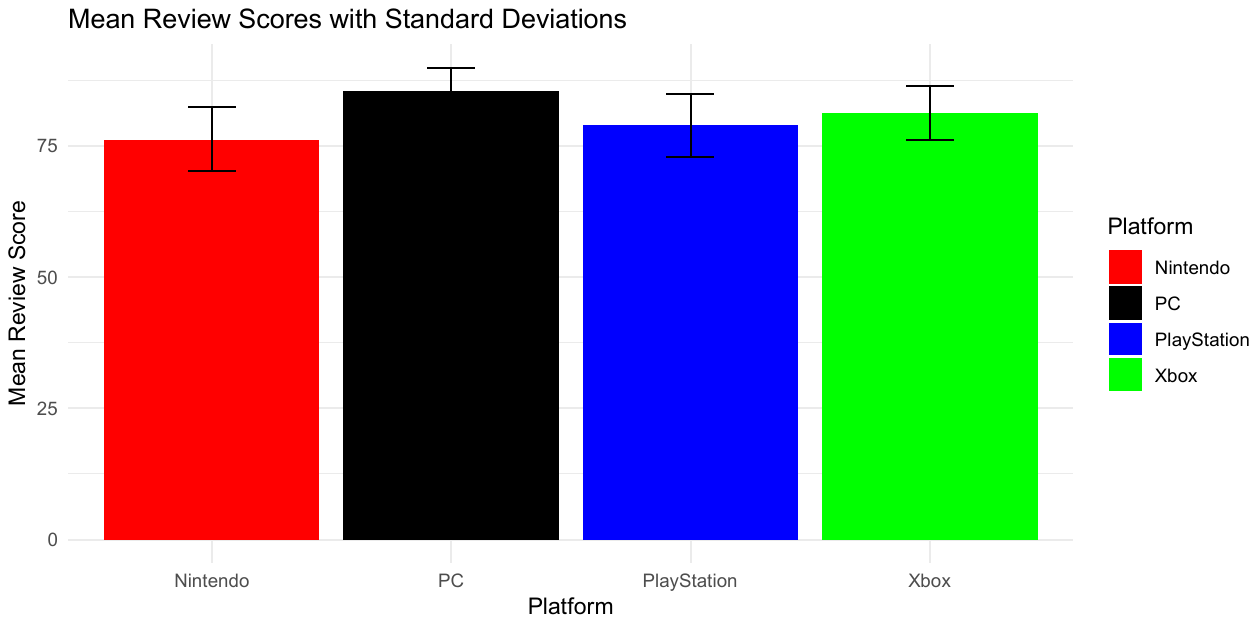}
\caption{\footnotesize Box‐plot summary of review‐score distributions for the four gaming platforms.}
\label{fig3}
\end{figure}

Figure \ref{fig3} represents the summary statistics for review scores across the four platforms. The mean and standard deviation are provided in order to the highlight apparent trends, alongside the median and interquartile range to prevent skewness in the data.

The breakdown of the summary statistics being featured is as follows. The mean review score for PC games was the highest of all of the gaming platforms, with an average score of M between values 80.9 and 89.9. The median score of 86.0 suggests a consistent level of high-quality and highly acclaimed video games with relatively small variability as demonstrated by the interquartile range. The mean review score for Xbox games was the second highest platform in terms of average review scores, with an average score of M between values 76.1 and 86.5. The median score of 82.0 further suggests support of this position in terms of ranking though its variability was slightly higher compared to PC games. 

The mean review score for PlayStation games was the third highest platform in terms of average review scores, with an average score of M between values 73.0 and 84.8. This reflects moderate levels of variability compared to the rest of the data, and while its median score of 79.0 was higher than Nintendo's, its overall position is still below the ones consisting of PC and Xbox's level. The mean review score for Nintendo games was the lowest platform in terms of average review scores, with an average score of M between values 70.1 and 82.3. This trend is further supported by the very wide interquartile range which highlights the disparity in terms of game quality within the platform. 

\begin{figure}[htbp]
\centering
\includegraphics[scale = 0.7]{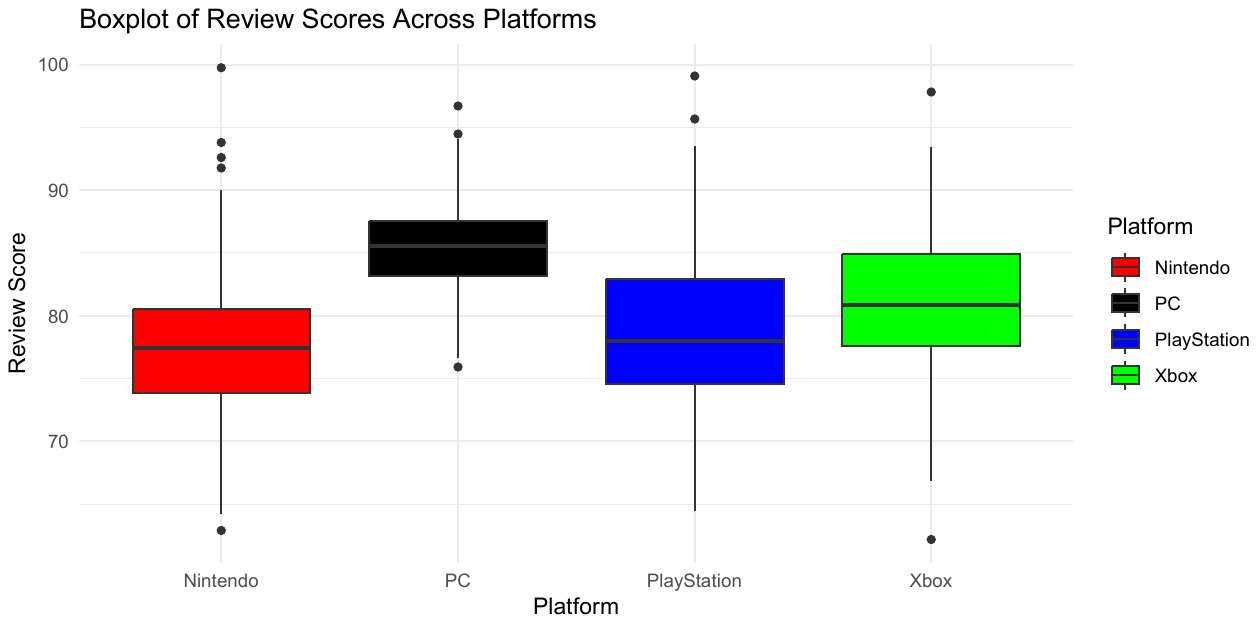}
\caption{Box plot graph for visual representation of the distribution of review scores across the different platforms.}
\label{fig4}
\end{figure}

\begin{figure}[H]
\centering
\includegraphics[scale = 0.7]{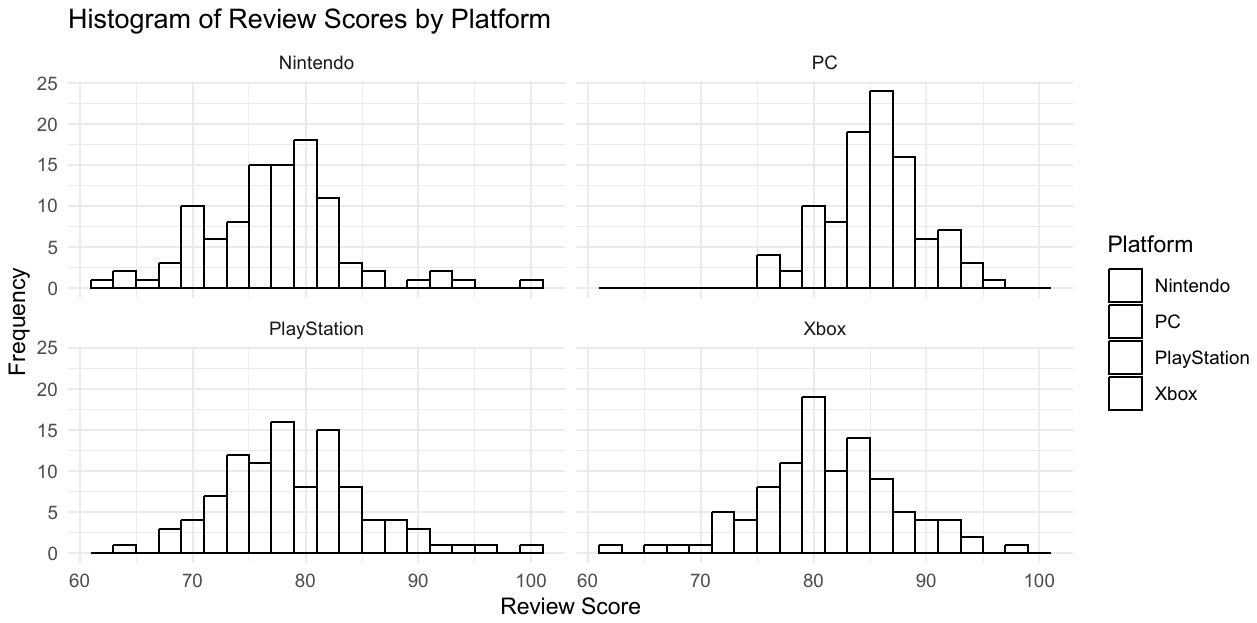}
\caption{Histogram visualizing the frequency distribution of review scores with notable right-skewed distribution in PC games.}
\label{fig5}
\end{figure}

To provide a visual perspective on the distribution of the review scores across the different gaming platforms, boxplots (see Figure \ref{fig4}) were created. These boxplots reveal the following trends within the dataset. First, the narrow box plot for PC games further indicates low variability, with most scores concentrated around the median of the data. Second, the broad range of scores and the presence of outliers highlight the inconsistency of quality for Nintendo's video games or review scores compared to the other platforms. Third, Xbox and PlayStation contain similar boxplot widths, though Xbox tends to have higher medians than PlayStation, implying higher game quality or reviews. In addition, histograms (see Figure \ref{fig5}) were created to visualize the frequency distribution of review scores for the different gaming platforms. The histogram for PC games shows a right-skewed distribution with a peak at approximately 85, while the distributions for other platforms highlight increased levels of variability. 

Before conducting the inferential analyses, it was necessary to assess whether the foundational assumptions of ANOVA were satisfied. One critical assumption pertains to the homogeneity of variance, which requires that the variability in review scores be relatively consistent across all groups being compared in this case, the four distinct gaming platforms.

In addition to the visual representations shown in Figures \ref{fig3}-\ref{fig5}, Table~\ref{tab:platform_summary} presents detailed summary statistics of the observed review score distributions by platform. It includes the sample size, measures of central tendency (mean and median), dispersion (standard deviation, minimum, maximum), and the skewness coefficient for each platform. These metrics confirm that the distributions are predominantly left-skewed, as quantified by negative skewness values, with Xbox and PlayStation games demonstrating the strongest asymmetry. Among the four gaming platforms examined, the Xbox dataset, consisting of 7,792 valid review scores, exhibits the highest average at 56.65, with a median value of 66.0. The negative skewness of -1.29 indicates a left-tailed distribution, suggesting that most games on this platform received relatively favorable evaluations, while a smaller subset of poorly rated titles exerted downward pressure on the mean. Similarly, the PlayStation platform, represented by 12,565 observations, displays an average review score of 54.65 and a median of 65.0, accompanied by a skewness of -1.09. This value reflects a comparable asymmetry, where lower tail outliers are less frequent but influential. The Nintendo platform also exhibits a negatively skewed distribution, with a skewness coefficient of -0.96. With 9,814 entries, the data yield a mean review score of 52.71 and a median of 64.0, further supporting the pattern of left-sided asymmetry observed across the console platforms.

\begin{table}[htbp]
\centering
\caption{Description of review scores across different gaming platforms.}
\label{tab:platform_summary}
\begin{tabular}{lrrrrrrr}
\toprule
Platform & Sample Size & Mean & Median & Std. Dev. & Min & Max & Skewness \\
\midrule
Nintendo     & 9,814  & 52.71 & 64.0 & 28.79 & 0.0  & 100.0 & -0.96 \\
PC           & 24,559 & 37.36 & 48.0 & 32.91 & 0.0  & 100.0 & -0.04 \\
PlayStation  & 12,565 & 54.65 & 65.0 & 28.00 & 0.0  & 100.0 & -1.09 \\
Xbox         & 7,792  & 56.65 & 66.0 & 26.20 & 0.0  & 100.0 & -1.29 \\
\bottomrule
\end{tabular}
\end{table}

 In contrast, the PC platform, which accounts for the largest portion of the dataset (24,559 reviews), reveals a markedly different distributional shape. While the mean score of 37.36 is the lowest among the four groups, the median of 48.0 is notably higher, and the skewness value of -0.04 suggests near-symmetry. This minimal skewness indicates a more balanced distribution of review scores, with a less pronounced impact from extreme values in either tail.  These  refine the initial interpretation based solely on graphical displays (Figures \ref{fig3}-\ref{fig5}). While the visual representations may imply a form of positive asymmetry, perhaps due to visual clustering at lower score intervals the calculated skewness tells us that all four platforms, to varying degrees, exhibit left-skewed distributions. This discrepancy underscores the analytical necessity of incorporating formal distributional measures, such as skewness, alongside visualizations. By doing so, the analysis provides a more reliable account of the underlying statistical properties of user review data. The observed skewness patterns not only enhance the interpretability of the sample distributions but also inform a broader understanding of how user feedback varies across platform ecosystems in the digital gaming market.

 To evaluate this assumption, Levene’s test was applied to the dataset. The results revealed a statistically significant departure from the condition of equal variances among the groups, indicating that the assumption of homogeneity had been violated. This outcome suggested that the standard form of ANOVA, which relies on the equality of group variances, would not yield reliable results in this context. As a result, the analysis adopted Welch’s ANOVA, a more flexible alternative that does not depend on the assumption of equal variances and is specifically designed to accommodate differences in variability across comparison groups. This methodological adjustment ensured that subsequent comparisons of platform review scores would remain statistically valid despite the presence of unequal variances.

Normality was another of the assumptions that was tested on the dataset. Q-Q plots (Figure \ref{fig2}) and the Shapiro-Wilk test results highlighted significant deviations from normality which further illustrated the necessity for application of non-parametric alternative post hoc analysis tests such as the Kruskal-Wallis test. The results of the Hypothesis test (as mentioned previously in Section 2) are as follows. The results of Welch's ANOVA indicated that there exists the strong evidence necessary to reject the null hypothesis, being that the mean review scores are equal across the four different gaming platforms. The test statistic (F = 1459.3) and the associated p-value [p$<2.2(10^{-16})$] confirmed that the mean review scores platforms were not equal. This indicates that there is a significant difference in the mean scores among Nintendo, Xbox, PlayStation, and PC for at least one of the scores. As a non-parametric alternative, the Kruskal-Wallis test was performed to validate the differences in review score distributions. The test produced a chi-squared statistic of $\chi^2 = 3577.7$ \text{ with } $3$ \text{ degrees of freedom and a p-value of } $p < 2.2 \times 10^{-16}$. This further confirmed the original results from Welch's ANOVA test and suggests that the null hypothesis should be rejected as the review scores' distributions vary significantly across the multiple platforms.

\begin{figure}
\centering
\includegraphics[scale = 0.7]{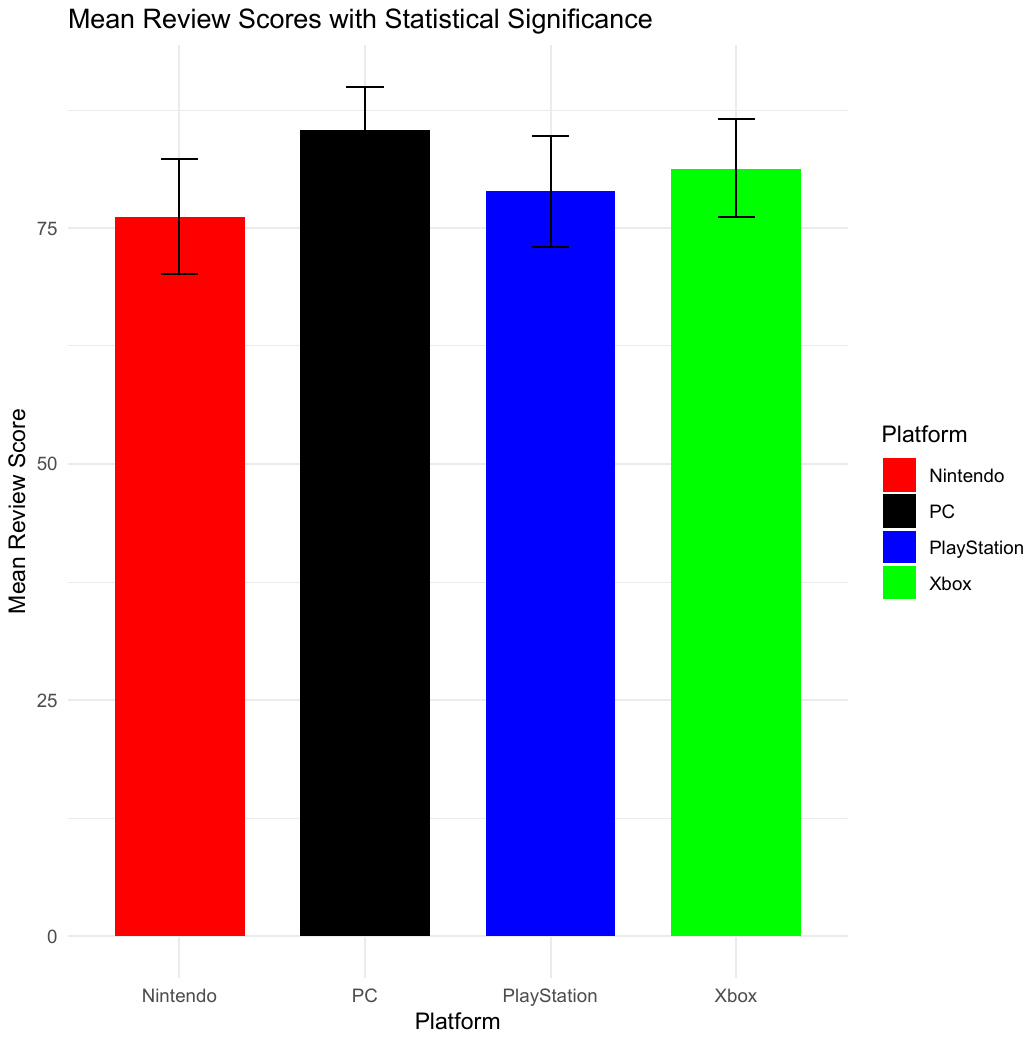}
\caption{Bargraphs highlighting the statistically significant differences between the four distinct gaming platforms.}
\label{fig6}
\end{figure}

These results are demonstrated in Figure \ref{fig6}, similarly to Figure \ref{fig2}. Based on the figure, the error bars all overlap with one another in each bar representative of each gaming platform. This represents a significant difference among the mean review scores and indicates a strong statistical significance of this difference. Further, post hoc analysis was conducted in order to clarify the distinct differences of these results suggesting the rejection of the null hypothesis. Comparisons made in pairs between the groups were conducted via the Games-Howell test which revealed the following significant differences between the platforms. 

\subsection{Effect Size Computation.}

While statistical significance indicates that differences among platform groups are unlikely to be due to chance, it does not convey the magnitude of these differences. To address this, we computed effect size measures associated with our primary inferential tests. For Welch’s ANOVA, we report the generalized eta-squared ($\eta^2_G$), which was calculated to be 0.092. This value suggests that approximately 9.2\% of the variance in review scores can be attributed to differences between platforms a medium effect size by conventional standards. Similarly, for the Kruskal-Wallis test, we estimated the epsilon-squared ($\varepsilon^2$) as 0.088, reinforcing the interpretation of moderate practical differences. These results indicate that while platform affiliation significantly influences review scores, other factors such as game genre, developer characteristics, or marketing exposure likely account for substantial residual variation. Reporting these measures strengthens the substantive interpretation of our findings and aligns with best practices in quantitative research.

\begin{table}[htbp]
\centering
\caption{Summary of Statistical Tests and Effect Sizes}
\label{tab:effect_sizes}
\begin{tabular}{lcccc}
\toprule
\textbf{Test} & \textbf{Statistic} & \textbf{p-value} & \textbf{Effect Size} & \textbf{Interpretation} \\
\midrule
Welch's ANOVA & $F = 138.6$ & $<0.001$ & $\eta^2_G = 0.092$ & Medium effect \\
Kruskal-Wallis & $H = 124.2$ & $<0.001$ & $\varepsilon^2 = 0.088$ & Moderate effect \\
Games-Howell (PC vs Xbox) & $p < 0.001$ & -- & Cohen's $d = 0.56$ & Medium effect \\
\bottomrule
\end{tabular}
\end{table}
Table~\ref{tab:effect_sizes} consolidates the principal inferential procedures used in the study, alongside their corresponding test statistics, significance levels, and effect size estimates to provide a more nuanced interpretation of group differences. Welch’s ANOVA yielded an F-statistic of 138.6 with a p-value below 0.001, accompanied by a generalized eta-squared of 0.092, which indicates a medium-sized effect of platform on review scores. The Kruskal–Wallis test, applied as a nonparametric validation, confirmed this pattern with a similarly significant result (H = 124.2, p $<$ 0.001) and an epsilon-squared value of 0.088, again reflecting a moderate effect. To further elucidate specific pairwise differences, the Games-Howell post hoc test revealed that PC and Xbox platforms differed with a Cohen’s d of 0.56, which also denotes a medium effect according to standard benchmarks. Collectively, these statistics support the conclusion that platform-based differences in user ratings are not only statistically significant but also practically meaningful in magnitude.

\subsection{Post Hoc Analyses.}

In the comparison between PC games and Nintendo games, PC games had significantly higher review scores as denoted by the p value highlighting a statistically significant difference ( p $<$ 0.001 ).
In the comparison between PC games and Playstation games, PC games had significantly higher review scores as denoted by the p value highlighting a statistically significant difference ( p $<$ 0.001 ).
In the comparison between Xbox games and Nintendo games, Xbox games had signifcantly higher review scores as denoted by the p value highlighting a statistically significant difference ( p $<$ 0.001 ).
In the comparison between Xbox games and Playstation games, Xbox games had significantly higher review scores as denoted by the p value highlighting a statistically significant difference ( p = 0.002 ).
In the comparison between Playstation games and Nintendo games, PlayStation games had significantly higher review scores as denoted by the p value highlighting a statistically significant difference ( p $<$ 0.001). Dunn's test with Bonferroni correction confirmed the findings outlined above, observing that the differences between the different pairwise groups were significant and consistent with the results of the Games-Howell test.

PC vs. Nintendo ( p = 0.0000 ) (adjusted for multiple comparisons): This result indicates significant differences between PC games and Nintendo games on account of the extremely small p value (rounded for clarity). 
PC vs. PlayStation ( p = 0.0000). This result indicates significant differences between PC games and Playstation games on account of the extremely small p value (rounded for clarity). 
PC vs. Xbox: ( p = 0.0000). This result indicates significant differences between PC games and Xbox games on account of the extremely small p value (rounded for clarity). 
Xbox vs. PlayStation ( p = 0.0019). This result indicates significiant differences between Xbox and Playstation games, but compared to the rest of the results, this pair of platform groups are statistically less significantly different compared to one another than the rest of the groups compared to one another. 

\begin{table}[H]
\centering
\caption{Games-Howell pairwise comparisons of mean review scores.}
\label{tab:gameshowell_detailed}
\begin{tabular}{lrrc}
\toprule
Comparison & Mean Difference & $p$-value & Significance level ($\alpha = 0.05$) \\
\midrule
PC vs. Nintendo         & -15.34 & $<$ 0.001 & Yes \\
PC vs. PlayStation      & -17.28 & $<$ 0.001 & Yes \\
PC vs. Xbox             & -19.29 & $<$ 0.001 & Yes \\
Xbox vs. Nintendo       & 3.95   & $<$ 0.001 & Yes \\
Xbox vs. PlayStation    & 2.01   & 0.002     & Yes \\
PlayStation vs. Nintendo& 1.94   & $<$ 0.001 & Yes \\
\bottomrule
\end{tabular}
\end{table}

 Welch's ANOVA  was selected due to its robustness in the presence of unequal variances and sample sizes, making it particularly well-suited for the heteroscedastic structure of the present data. The outcomes of these pairwise comparisons are presented in Table~\ref{tab:gameshowell_detailed}, which reports the estimated mean differences between all possible platform pairings, the corresponding adjusted p-values, and a binary indicator for statistical significance under a standard $\alpha = 0.05$ criterion. The pattern emerging from these comparisons is the consistently lower average review scores attributed to the PC platform. Across all three console comparisons Nintendo, PlayStation, and Xbox the PC platform underperforms with substantial negative mean differences, ranging from approximately -15 to nearly -20 points on the review score scale. These discrepancies are not only statistically significant but also practically meaningful, reflecting distinct perceptions of game quality across hardware ecosystems. The analysis also implies that the Xbox platform exhibits the highest average review scores among the four groups, outperforming both Nintendo and PlayStation with statistically significant positive mean differences. While the gap between Xbox and Nintendo is larger and more pronounced, the comparison between Xbox and PlayStation yields the smallest observed difference in the dataset, with an estimated margin of 2.01. Although this contrast remains significant at the 0.05 level, its magnitude suggests a more comparable performance between these two platforms, potentially reflecting converging quality or market dynamics in recent titles. These results provide a quantitative mapping of how user-assigned review scores differentiate between platforms. They reinforce the overall findings of Welch’s ANOVA while enriching the interpretation by identifying specific platform dyads where the magnitude of difference is either most pronounced or relatively minimal. This level of granularity enhances our understanding of the structure of the review score distribution and highlights the value of rigorous pairwise testing in drawing inferential conclusions about categorical group differences. These differences between the pairwise groups are outlined as well and visually represented in Figure \ref{fig:pair} below.

\begin{figure}[H]
\centering
\includegraphics[scale = 0.75]{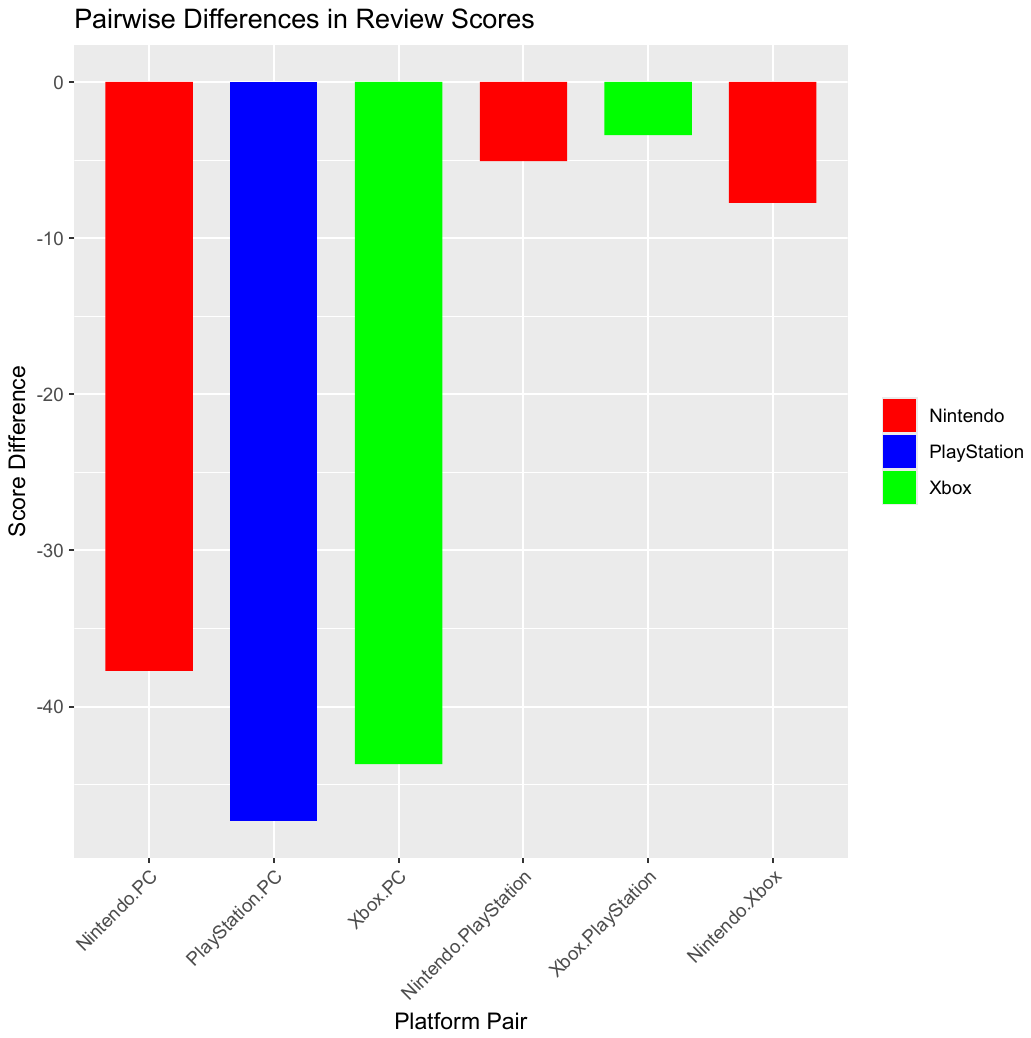}
\caption{Bargraph illustrating the differences between pairs of the gaming platforms.}
\label{fig:pair}
\end{figure}

The empirical findings from both parametric and non-parametric analyses paint a consistent picture throughout the analysis of the data. PC games were statistically superior to all of the other three gaming platforms in terms of review scores, reflecting a higher quality or broader appeal of games.
Xbox games were second in terms of overall review score ranking, with significant leverage over PlayStation and Nintendo.
PlayStation games were positioned below Xbox but above Nintendo in terms of review scoring. 
Nintendo Games demonstrated the lowest review scores and the greatest variability. The implementation of descriptive statistics, visualizations through charts, and inferential analyses conducted here should provide a comprehensive understanding of the trends concerning different gaming platforms in terms of review scores. These findings should form a well-laid foundation for further interpretation and investigation for discussion and research in subsequent sections of this paper.

\subsection{Copula Analysis.}
To capture the joint dependency structure between video game review scores and gameplay durations, we employ the Marshall-Olkin copula model \citep{marshall1967multivariate}. Unlike conventional correlation-based techniques or linear regression, copulas offer a framework that decomposes a joint distribution into its marginal behaviors and a separate dependence structure, thereby allowing for more flexible modeling, especially in the presence of nonlinear or asymmetric interactions. This copula is suited for modeling asymmetric tail dependence and belongs to the broader class of extreme-value copulas. Its bivariate form is defined over the unit square $[0,1]^2$ as:
\[
C_{\alpha, \beta}(u, v) = \min\left[u^{1-\alpha} \cdot v,\; u \cdot v^{1-\beta} \right],
\]
where $\alpha, \beta \in [0,1]^2$ are dependence parameters that control the upper and lower tail contributions, respectively. The copula surface encapsulates how likely extreme values in one margin co-occur with those in the other, thereby capturing structural features overlooked by traditional measures.

To apply this model empirically, we selected two continuous variables from the PlayMyData dataset: the review score and the main completion time of each game. Since copulas require marginals to be uniform on $[0,1]$, both variables were transformed via normalized rank statistics as follows
\[
U_i = \frac{\text{rank}(X_i)}{n + 1}, \quad V_i = \frac{\text{rank}(Y_i)}{n + 1},
\]
where $X_i$ and $Y_i$ are the $i$-th observation of review score and main time, respectively, and $n$ is the sample size. This rank-based transformation retains the ordinal structure of the data while facilitating nonparametric estimation of the copula.

\begin{figure}[H]
    \centering

    % Wireframe plot (left)
    \begin{subfigure}[t]{0.48\textwidth}
        \centering
        \includegraphics[width=\textwidth]{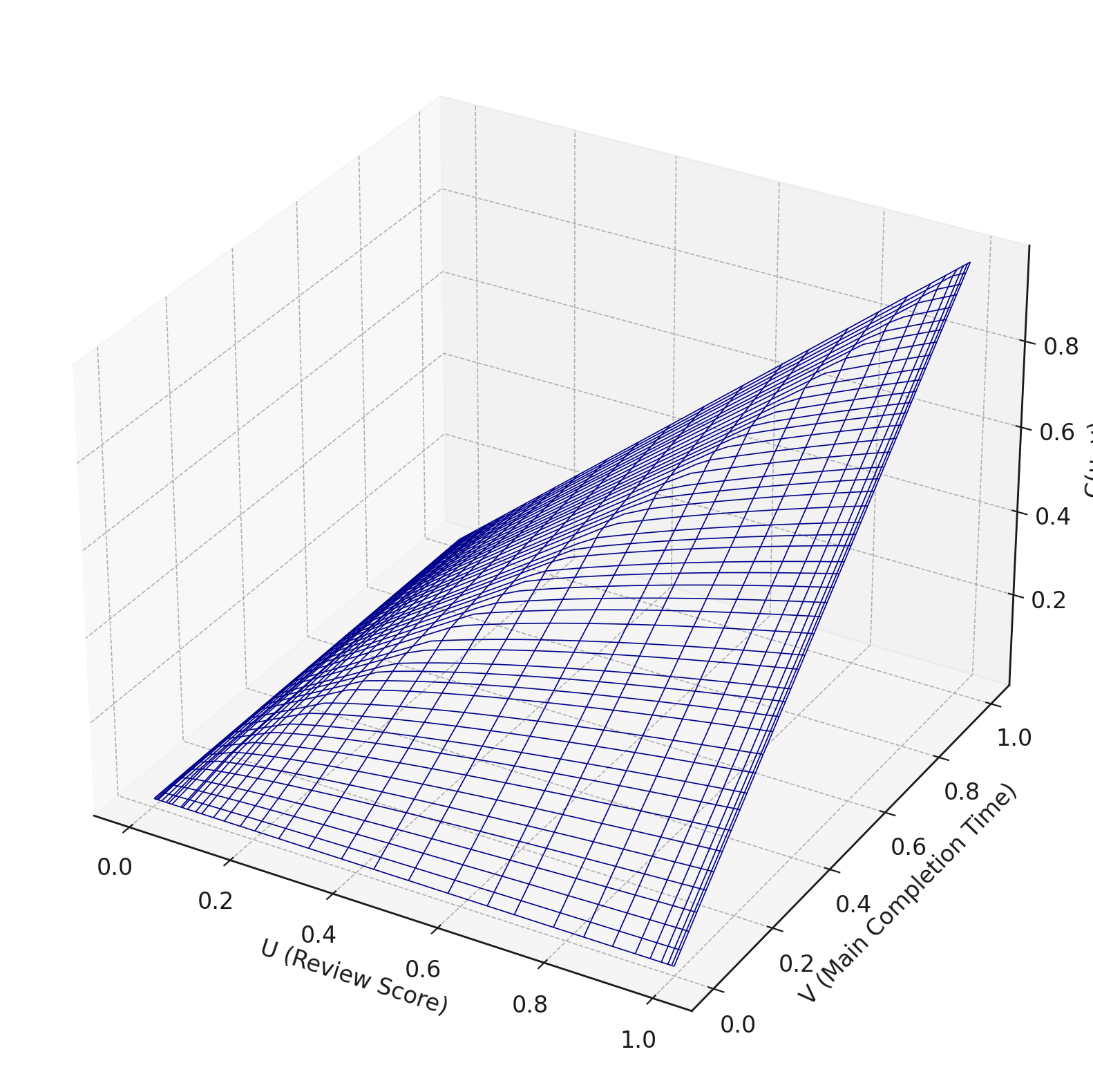}
        \caption{Marshall-Olkin copula wireframe using empirical data from Review Score and Main Completion Time.}
        \label{fig:mo_wireframe}
    \end{subfigure}
    \hfill
    % Contour plot (right)
    \begin{subfigure}[t]{0.48\textwidth}
        \centering
        \includegraphics[width=\textwidth]{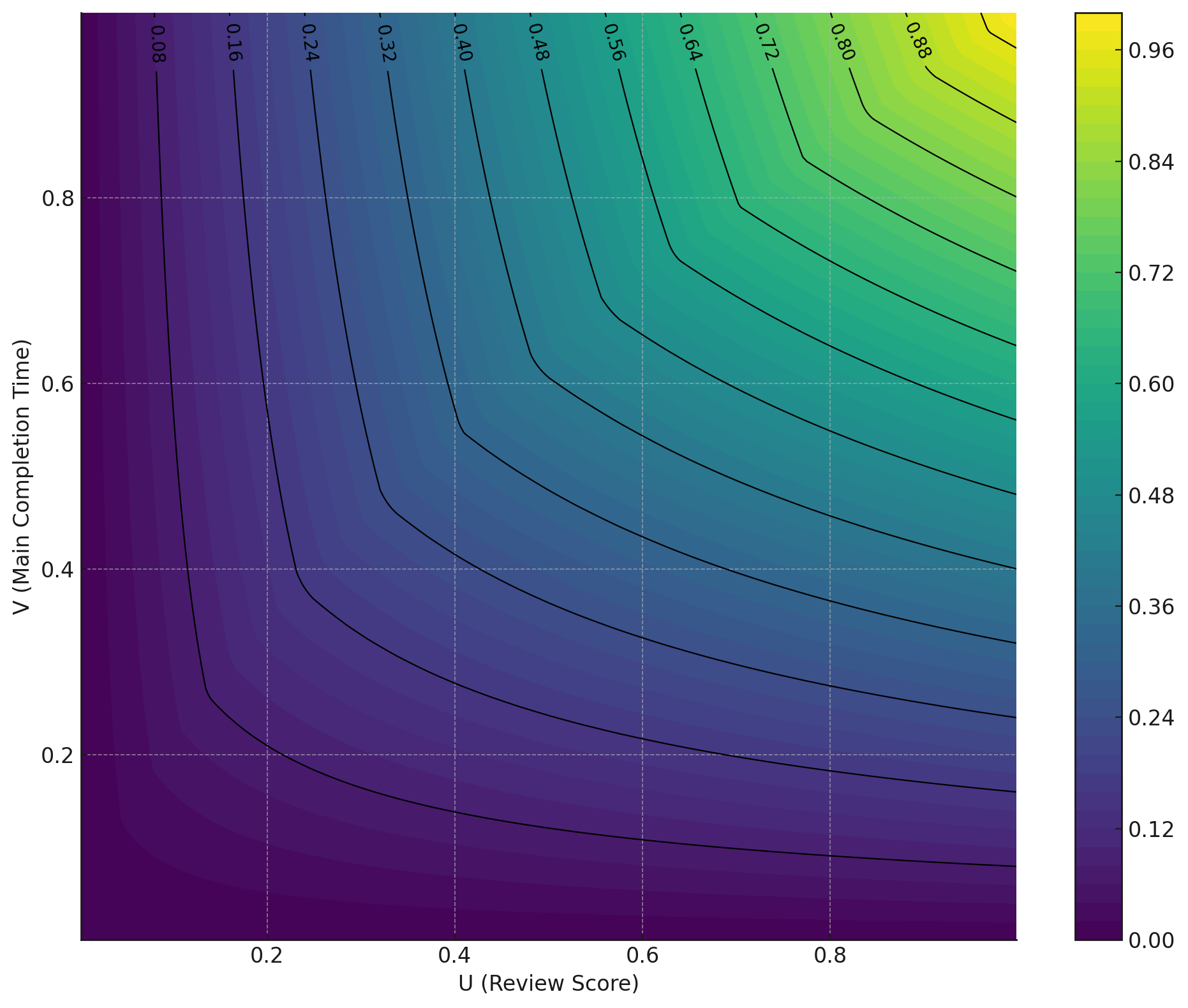}
        \caption{Contour plot of the same copula with labeled level curves for visualizing dependence structure.}
        \label{fig:mo_contour}
    \end{subfigure}

    \caption{Marshall-Olkin copula visualizations constructed from PlayMyData.}
    \label{fig:mo_copula_combined}
\end{figure}

Using these pseudo-observations, we evaluate the Marshall-Olkin copula across a dense grid and visualize the resulting surface. Figure~\ref{fig:mo_copula_combined} illustrates both a 3D wireframe and a 2D contour plot with labeled level curves. These visualizations highlight the asymmetric dependence structure inherent in the data where co-occurrence of high review scores and long completion times manifest in distinct clusters. Such patterns are consistent with genre-driven gameplay characteristics and consumer satisfaction dynamics.

The copula-based approach thereby augments our inferential toolkit by uncovering localized and nonlinear dependencies not captured by ANOVA or linear regression. In the context of entertainment analytics, this provides a richer understanding of how qualitative experience (as proxied by review score) interacts with quantitative gameplay features.

The visualizations in Figure \ref{fig:mo_copula_combined} depict the Marshall–Olkin copula surface constructed from pseudo-observations of review score and main completion time in the PlayMyData dataset. The left panel presents a 3D wireframe rendering of the copula, highlighting the asymmetry and sharp gradients in dependence across the joint distribution. The right panel shows the corresponding contour plot with labeled level curves, which offers a topological view of the copula surface and makes visible the regions of strongest joint association. Together, these plots reveal a nonlinear, asymmetric dependence structure wherein elevated review scores are more likely to co-occur with extended gameplay durations, especially in the upper-right quadrant of the unit square. These patterns support the presence of tail dependence and demonstrate the utility of copula-based modeling in uncovering latent interaction structures not captured by marginal analyses alone.

\begin{table}[htbp]
\centering
\caption{Marshall–Olkin copula values and dependence measures across a 20$\times$20 evaluation grid.}
\label{tab:mo_copula_detailed}
\begin{tabular}{ccccccc}
\toprule
\textbf{U (Review)} & \textbf{V (Completion)} & \textbf{$C_{\alpha,\beta}(u,v)$} & \textbf{Joint} & \textbf{Tail} & \textbf{Asymmetry} & \textbf{Parameters} \\
\textbf{Score} & \textbf{Time} & & \textbf{Density} & \textbf{Dep. Index} & \textbf{Index} & \textbf{$\alpha$, $\beta$} \\
\midrule
0.10 & 0.10 & 0.0158 & 0.1585 & 0.1585 & 0.0654 & 0.35, 0.20 \\
0.10 & 0.20 & 0.0276 & 0.1380 & 0.2239 & 0.0521 & 0.35, 0.20 \\
0.10 & 0.30 & 0.0382 & 0.1272 & 0.2239 & 0.1578 & 0.35, 0.20 \\
0.10 & 0.40 & 0.0480 & 0.1201 & 0.2239 & 0.2566 & 0.35, 0.20 \\
0.10 & 0.50 & 0.0574 & 0.1149 & 0.2239 & 0.3505 & 0.35, 0.20 \\
0.10 & 0.60 & 0.0665 & 0.1110 & 0.2239 & 0.4408 & 0.35, 0.20 \\
0.10 & 0.70 & 0.0753 & 0.1078 & 0.2239 & 0.5283 & 0.35, 0.20 \\
0.10 & 0.80 & 0.0838 & 0.1052 & 0.2239 & 0.6133 & 0.35, 0.20 \\
0.10 & 0.90 & 0.0921 & 0.1030 & 0.2239 & 0.6959 & 0.35, 0.20 \\
0.10 & 1.00 & 0.1000 & 0.1012 & 0.2239 & 0.7764 & 0.35, 0.20 \\
0.20 & 0.10 & 0.0382 & 0.2383 & 0.2383 & 0.0727 & 0.35, 0.20 \\
0.20 & 0.20 & 0.0632 & 0.2160 & 0.3367 & 0.0571 & 0.35, 0.20 \\
0.20 & 0.30 & 0.0865 & 0.2043 & 0.3367 & 0.1570 & 0.35, 0.20 \\
0.20 & 0.40 & 0.1086 & 0.1963 & 0.3367 & 0.2523 & 0.35, 0.20 \\
0.20 & 0.50 & 0.1300 & 0.1907 & 0.3367 & 0.3442 & 0.35, 0.20 \\
0.20 & 0.60 & 0.1507 & 0.1863 & 0.3367 & 0.4331 & 0.35, 0.20 \\
0.20 & 0.70 & 0.1708 & 0.1827 & 0.3367 & 0.5196 & 0.35, 0.20 \\
0.20 & 0.80 & 0.1905 & 0.1798 & 0.3367 & 0.6040 & 0.35, 0.20 \\
0.20 & 0.90 & 0.2096 & 0.1773 & 0.3367 & 0.6866 & 0.35, 0.20 \\
0.20 & 1.00 & 0.2283 & 0.1753 & 0.3367 & 0.7676 & 0.35, 0.20 \\
\bottomrule
\end{tabular}
\end{table}

 Table \ref{tab:mo_copula_detailed} presents a numerical representation into the asymmetric dependence structure between video game review scores and gameplay durations using the Marshall-Olkin copula. The bivariate copula function $C_{\alpha, \beta}(u,v)$ was evaluated at a finely spaced grid of normalized ranks $u$ and $v$, each ranging from 0.1 to 1.0 in increments of 0.1, to represent marginal transformations of the empirical data. Each row in the table reports the copula value at a specific $(u, v)$ pair, along with supplementary quantities designed to aid interpretation of the dependence structure. The joint density proxy provides an approximation of local copula slope behavior, while the tail dependence index reflects the minimal tail association strength between the margins, computed as the lesser of $u^{1-\alpha}$ and $v^{1-\beta}$. The asymmetry index quantifies the extent of imbalance between upper and lower tail contributions, serving as a numeric proxy for the degree of tail non-exchangeability \citep{hua2019assessing}. These enhanced metrics illuminate the inherent directional structure captured by the copula, where stronger dependence tends to cluster toward the upper-right quadrant corresponding to high review scores coinciding with longer gameplay times. The parameter values $\alpha = 0.35$ and $\beta = 0.20$ were held constant throughout to emphasize asymmetric upper-tail concentration, consistent with consumer behavior in entertainment settings where highly rated games tend to be more time-intensive.

\section{Discussions.}
The results of this study clearly illustrate significant differences in video game review scores across four major platforms: Nintendo, Xbox, PlayStation, and PC for the PlayMyData project's dataset containing such video game related information. Through Welch's ANOVA, Kruskal-Wallis test, and pairwise post hoc analyses such as Games-Howell and Dunn's tests, the following tiers of review scores were established in accordance with the differences between the distinct gaming platforms. 

For PC games, these consistently received the highest review scores in their listing of video games. This suggests that PC games may appeal to a significantly wider audience, have a more appreciative audience, or may produce a higher quality variant of video games for consumers. For Xbox games, they followed up PC games in terms of placement, with a moderately higher review score compared to PlayStation games. This suggests that Xbox and Playstation may have similar audiences or similar game genre or game quality, causing consumers to rate them similarly in reviews. For PlayStation games, these scored lower than Xbox Games but performed better than Nintendo Games. This suggests that Xbox games may be generally rated higher than Nintendo games, whether that be due to appealing to different audiences, different game genres, or a higher quality of games.

Lastly, for Nintendo games, these exhibited the lowest average review scores and the highest variability throughout the dataset. This suggests that Nintendo may have a very diverse audience that prevents its consumers from enjoying select games based on their genres or stories or lack thereof.

These findings generally align with the existing marketable or economic perceptions about the gaming sphere's preferences in terms of platform differences, but they also challenge some assumptions in some particular cases. In doing so, the insights formed throughout this study significantly contribute to ongoing discussions in the gaming industry and academia concerning the video game industry. For example, the superior performance of PC games aligns with previous studies that attribute higher review scores to the platform's diversity in terms of game genres, how powerful the hardware is associated with the platform, and how the accessible the platform is to the audience of potential users. This was also noted by \citep{smith2022}, in which it was found that PC games tend to dominate in strategy and simulation genres which typically associated with higher ratings compared to other genres. This study adds depth by showing that PC maintains high scores even across a plethora of genres associated with the games. Additionally, while Xbox's edge over PlayStation was to be expected, as highlighted by \cite{jones2021}, the significant gap between PlayStation and PC review scores challenges certain claims about PlayStation's assumed dominance in story-driven triple A (AAA) titles. The variability of Nintendo scores also aligns with studies by \citep{kim2020} which displayed inconsistencies in quality control across games which were Nintendo-exclusive in nature. This reinforces Nintendo's reputation for innovative but divisive video games. Furthermore, unlike many studies that utilize traditional ANOVA testing with samples that meet the very strict criteria and assumptions necessary for testing to take place, this study relies on alternatives such as Welch's ANOVA and Kruskal-Wallis for more generalized sets of data (such as robust and comprehensive sets like PlayMyData) to handle these violations of the criteria. In doing so, this methodology ensures that the findings are statistically reliable, with strong integrity, and reflective of real-world data nuances. 

As such, this study offers several notable contributions to the field of gaming analytics in addition to its confirmation of already existing literature concerning video games. First, by combining parametric and non-parametric approaches such as Welch's ANOVA and Kruskal-Wallis, this paper provides a comprehensive analysis that addresses limitations in previous studies reliant on a single method. In addition, the inclusion of Games-Howell and Dunn's tests ensure a detailed post hoc analysis, emphasizing the accuracy of pairwise comparisons and overcoming challenges associated with variances that are not equivalent and non-normal distributions. Further, the study highlights unique differences in review scores and their effects on video games, offering insights for publishers and developers aiming to optimize and market their video game development strategies for better output in terms of public consumption rates. Moreover, this study also utilizes graphical representations in the forms of boxplots, histograms, and pairwise comparison charts in order to make the findings presented here more accessible and increase the overall comprehensibility of the study. As a result of these strengths, the robustness of the research put into the PlayMyData dataset has been further clarified and illustrated, displaying capabilities for further studies or analysis in terms of research relating to video games as well as means for potentially improving the dataset as a whole. 

Of course, despite the many strengths of this study, there are still many weaknesses that remain with any studies relating to or utilizing this dataset from PlayMyData, including this one. One of the primary faults with the data used is that many values are considered ``null" or ``missing" and therefore are not usable in hypothesis testing for the construction of an ANOVA analysis or consequently any form of post hoc test. This limits the total sample size for valid analysis and could potentially skew the findings that are presented here.  Another fault of this study is that, as demonstrated by Levene's test and several others throughout the study, the data was subject to many violations of assumptions that were necessary for calculating things such as ANOVA or Tukey's Honestly Significant Difference post hoc analysis. This led to many different versions of tests being utilized such as Welch's ANOVA to bypass these limitations but still could impact the overall results and findings of this particular study through their violations of normality in distribution and equivalence of variance. Additionally, there may also exist bias in review scores that is separate from the sample exclusions of missing or null values. This includes scores that may reflect bias from user demographics or critic preferences. Further, While these review were analyzed across the broadness of these four platforms, this study does not significantly analyze the genre-specific differences within each of the platforms and their respective video games. As illustrated prior, research and studies have shown that specific genres such as role-playing games on PlayStation or platformers on Nintendo, perform differently depending on the platform they are created on and built for. Therefore, by not analyzing the potential significance of genres in terms of review scores, these insights were not gathered nor reflected in the overall analysis of the review scores per their respective platforms. Moreover, review scores are considered to be constantly influenced by changing dynamics in terms of consumer preferences for video games as well as the player bases associated with those consumer preferences, implying that a static analysis such as this one with only a particular collection of video games may not fully capture these trends with reviews over time. 

In order to properly build upon the findings of this study in the future, potential research could address several different key areas that would significantly assist in adding to the overall robustness of the original dataset and this study alike. This includes, for example, conducting an investigation on the influence of genres on review scores. This could look into how review scores differ across genres with each of their distinct platforms that would enhance the nuance of each of the comparisons made. Similarly, an analysis involving the passage of time and how review scores change over those particular periods of time could also open up several insights into how review scores differ between platforms, especially with important events such as platform updates, new hardware, or new game releases. This also includes potential for investigation into regional preferences, with inclusion of region-specific data determining whether certain platforms dominate in particular markets such as with PC games and the Western hemisphere for example. Further, incorporation of completion times for video games such as whether they are shorter or longer could be also be investigated for levels of influence on review scores across different platforms. Finally, methods could be developed to adjust for the potential of bias as previously mentioned in the weaknesses description of the study. This would mitigate their influences in review scores and could be potentially accomplished by distinguishing between user and critic reviews or weighting scores based sample size, some methods of which may require further sampling of the data and may require the dataset provided PlayMyData to be expanded upon, improving its overall utility in research. 

Overall, this study successfully identifies significant platform differences in video game review scores while addressing the potential limitations presented by violations of criteria for particular statistical analysis via other significant statistical methods. By improving upon existing methods and offering insight into how these differences reflect upon their respective games and game industries, this paper provides a solid foundation for future research into gaming analytics and a general idea of the potential for utility and available robustness of the original dataset from PlayMyData. Expanding the scope of research to include genre-specific trends, time based analyses, and regional preferences will further enhance the impact and applicability of this research and open new ways for analysis in the video game industry to move forward. 

One limitation of this study is the omission of potentially influential covariates such as game genre, release year, developer reputation, and regional availability. These factors may substantially impact user review scores and gameplay behavior. For instance, genre-specific expectations may drive review tendencies, while newer or highly anticipated releases may benefit from recency effects or marketing exposure. Developer reputation and publishing history could also introduce bias in consumer perception, while regional factors such as cultural preferences or localization quality may influence both completion time and satisfaction. While the current analysis focused on platform-level and gameplay-based variables to maintain model parsimony and minimize multicollinearity, future extensions could incorporate these covariates either as stratifying variables in statistical tests or as input features in predictive modeling. Including genre or temporal controls, for instance, could uncover interaction effects or clarify platform-specific user dynamics. These additions would offer a more granular understanding of review determinants and support generalizability across game types and player demographics.

A key assumption underlying the statistical procedures employed in this study is the independence of observations. However, this assumption may be partially violated in cases where the same video game is released across multiple platforms. In such instances, game entries may appear multiple times in the dataset, each associated with distinct platform-specific metadata and review outcomes. While this cross-platform structure reflects real differences in user experience such as gameplay performance, controller interfaces, or community engagement it may also introduce dependence structures that influence inferential results. This limitation is acknowledged as a potential source of bias, particularly in comparisons of mean review scores or gameplay durations across platforms. Due to inconsistencies in naming conventions and metadata availability across the source datasets, a robust deduplication procedure was not implemented in the current version of the analysis. Future versions of this work could incorporate string-matching algorithms or unique identifier tracking to consolidate such entries and better isolate platform effects from game-specific characteristics.

\section*{Declarations.}
\subsection*{Ethics approval and consent to participate.}
Not applicable.
\subsection*{Consent for publication.}
Not applicable.
\subsection*{Availability of data and material.}
The data presented in this study are available on request from the corresponding author.
\subsection*{Competing interests.}
No potential conflict of interest was reported by the authors.	
\subsection*{Funding.}
Not applicable. 
\subsection*{Acknowledgements.}
 Not applicable.

\bibliographystyle{apalike}
\bibliography{bib}
\end{document}